\documentclass{aastex}
\usepackage{emulateapj5}

\newcommand\pbar{$\bar{p}\mbox{ }$}
\newcommand\dbar{$\bar{D}\mbox{ }$}
\newcommand\hebar{$\bar{He}\mbox{ }$}
\newcommand\xbar{$\bar{X}\mbox{ }$}

\begin{document}
\submitted{Accepted for publication in the Astrophysical Journal}
\title{A  novel antimatter  detector  based on  X-ray  deexcitation of  exotic
atoms}
\author{
Kaya Mori\altaffilmark{1}, 
Charles J. Hailey\altaffilmark{1},
Edward A. Baltz\altaffilmark{1}, 
William W. Craig\altaffilmark{2},
Marc Kamionkowski\altaffilmark{3},
William T. Serber\altaffilmark{1}, 
and Piero Ullio\altaffilmark{4}
}
\altaffiltext{1}{Columbia Astrophysics Laboratory, 538 W. 120th St., New York, NY 10027}
\altaffiltext{2}{Lawrence Livermore National Laboratory, 7000 East Avenue, Livermore CA, 94550}
\altaffiltext{3}{California Institute of Technology, Mail Code 130-33,
Pasadena, CA 91125}
\altaffiltext{4}{SISSA, via Beirut 4, S4014 Trieste, Italy}

\begin{abstract}

We  propose a  novel  antiparticle detector.  The gaseous  antiparticle
spectrometer (GAPS) effects particle identification through the characteristic
X-rays  emitted by  antiparticles  when they  form  exotic atoms  in
gases. GAPS  obtains particularly high grasp (effective area--solid angle
product) at lower particle energies, where conventional schemes are most
limited in their utility. The concept is simple and lightweight, so it can be
readily employed on balloon and space-based missions. An extremely powerful
potential application of GAPS is a space-based search for the neutralino
through the detection of a neutralino annihilation by-product -- the
antideuteron. Paradoxically, this space-based search for the neutralino is
capable of achieving comparable sensitivity to as yet unrealized third
generation, underground dark matter experiments. And GAPS can obtain this
performance in a very modest satellite experiment. GAPS can also provide
superior performance in searches for primary antiprotons produced via
neutralino annihilation, black hole evaporation and in probing subdominant
contributions to the antiproton flux at low energies. In a deep space mission
GAPS will obtain higher sensitivity for a given weight and power than BGO
calorimeters. 

\end{abstract}
\slugcomment{Accepted for publication in the Astrophysical Journal}


\section{Introduction}
\label{intro}

The gaseous antiparticle
spectrometer (GAPS) identifies antiparticles through the characteristic X-rays
emitted by antimatter when it forms exotic atoms in gases. GAPS provides more
than an order of magnitude or more improvement in sensitivity compared to
conventional magnetic spectrometers at substantially lower weight and
cost. GAPS is thus ideal for space-based experiments. In \S\ref{concept}  of the paper we describe the scientific opportunities
which can be exploited with the superior sensitivity of GAPS. We particularly
focus on a high sensitivity, indirect search for the neutralino through
detection of cosmic antideuterons. This approach to neutralino detection can
yield sensitivities comparable to, or even exceeding, those of as yet
unrealized third generation, underground dark matter experiments. We also
discuss more prosaic possibilities such as measurement of the very low energy
antiproton spectrum. We also mention more exotic possibilities such as
searching for antiprotons from evaporating black holes and searching for
antihelium with much greater sensitivities than the AMS experiment on
International Space Station (ISS). In \S2 we describe the basic GAPS
concept, in \S3 the atomic physics of exotic atoms, in \S4 the
detector efficiency, in \S5 issues of background rejection and in \S6 results
of preliminary simulation. 

\subsection{Indirect detection of dark matter through antideuterons}
\label{intro_dbar}

A major goal of 21st century physics is to identify particle dark matter. The
best candidate is likely a weakly interacting massive particle (WIMP)
\citep{jungman96} such as those that arise from supersymmetric extensions of
the standard model of particle physics. Most effort has concentrated on the
neutralino, the lightest supersymmetric partner.  The neutralino can be
detected by the nuclear recoils it produces through its scalar and vector
couplings to matter.  Major direct detection experiments are underway but they
are extremely difficult because of the low predicted count rate for the
neutralino. These experiments  must be done deep underground to shield against
cosmic-rays and in particular neutrons produced in the atmosphere and through
muon interactions (the tertiary neutrons).  Tertiary neutrons, as well as
those produced in ($\alpha,n$) reactions in the surrounding rock are
particularly problematic for these direct detection experiments. The
experiments cannot distinguish a target atom recoil due to a neutron from one
due to a neutralino. Therefore, Monte-Carlo simulations are required to
estimate neutron contribution to the nuclear recoil. The recent controversy over the possible
detection of the neutralino by DAMA and the contradictory claims of the CDMS
experiment hinge on the reliability of the neutron background estimation
\citep{akerib00}. This situation requires considerable caution and it has been
advocated that a reliable neutralino detection may require several experiments
operating with different target nuclei and obtaining consistent neutralino
detection rates. Ideally such experiments would have different intrinsic
background sources. However all underground direct detection experiments have
neutrons as the dominant source of uncertainty. 

Alternately, many indirect detection schemes have been proposed for the
neutralino \citep{gondolo00}. These rely  on  the fact  that  the  neutralino is  a
Majorana particle, and thus can annihilate with itself. The resultant heavy
quarks and gauge and Higgs bosons produce hadronic
and  electromagnetic showers. This leads to a primary antiproton component to
the cosmic-rays. The antiproton component has been discussed in many papers
and searched for in many experiments. It is difficult to distinguish the
primary antiproton component from the secondary component produced in
cosmic-ray interactions in the interstellar medium (\S\ref{intro_pbar}
below). Both line and continuum $\gamma$-ray signatures have been proposed as
a means to search for the neutralino with GLAST. In addition, it has been
proposed to search for the neutrino signature produced
when neutralinos annihilate in the gravitational potential well of the sun
\citep{amanda99}. 

The promise of indirect detection techniques to search for the neutralino has
changed markedly in the last year. Theoretical calculations predict a flux of
primary antideuterons in the cosmic-rays due to the annihilation of the
neutralino \citep{donato00}. Like the well-known primary antiproton signal, the
antideuteron signal is produced when WIMPs annihilate to heavy quarks and
gauge and Higgs bosons that fragment to cosmic-ray antiprotons and
antineutrons. This flux is large enough that the GAPS technique, when employed
in a modest space-based experiment, has competitive and possibly superior
sensitivity to as yet unrealized 3rd generation direct detection
experiments. Indeed the indirect detection of the neutralino via the
antideuteron provides an ideal complementary technique to the direct detection
experiments because its background source is not neutrons (see below). 

Figure \ref{sensitivity1} shows the projected sensitivity of CDMS II, a state-of-the-art
direct detection experiment, through the year 2004. Also shown is the
projected sensitivity of GENIUS, a proposed 3rd generation experiment based on
$\sim1$ Ton Germanium. The third generation experiments will improve on the
2nd generation experiments by about 3 orders of magnitude. The dots represent
the ensemble of SUSY models parameterized by their spin-independent
cross-section and neutralino mass. A similar plot is shown in figure \ref{sensitivity2}
for a modest MIDEX class satellite experiment (described in more detail in
\S\ref{app_dbar}). The sensitivity calculation is all-inclusive (including the
effects of orbit-varying geomagnetic cutoff and solar modulation). The
sensitivity for this (unoptimized) experiment is much more than an order of magnitude better
than AMS for detecting the antideuteron. Recently a g--2 experiment has
detected a marginal (2.7$\sigma$) discrepancy in the anomalous magnetic moment
for the muon \citep{brown01}. If this discrepancy is due to supersymmetric
corrections 
to loop diagrams producing the magnetic moment, then the range of possible
SUSY models is severely constrained, as shown by the green circles in figures
\ref{sensitivity1}, \ref{sensitivity2} \citep{baltz01}. Unlike the AMS 
experiment on ISS, the GAPS experiment on a small satellite can be as sensitive as a 3rd generation experiment and will access nearly the entire allowed SUSY parameter space as restricted by the g--2 results. 

\begin{figure}
\epsscale{0.5}

\caption{The interstellar flux of secondary antideuterons (heavier solid
curve) decreases at low energy whereas the energy spectrum of the
antideuterons from supersymmetric origin (curves a to d) tends to flatten
(from \citet{donato00}).\label{dbar_flux}}
\plotone{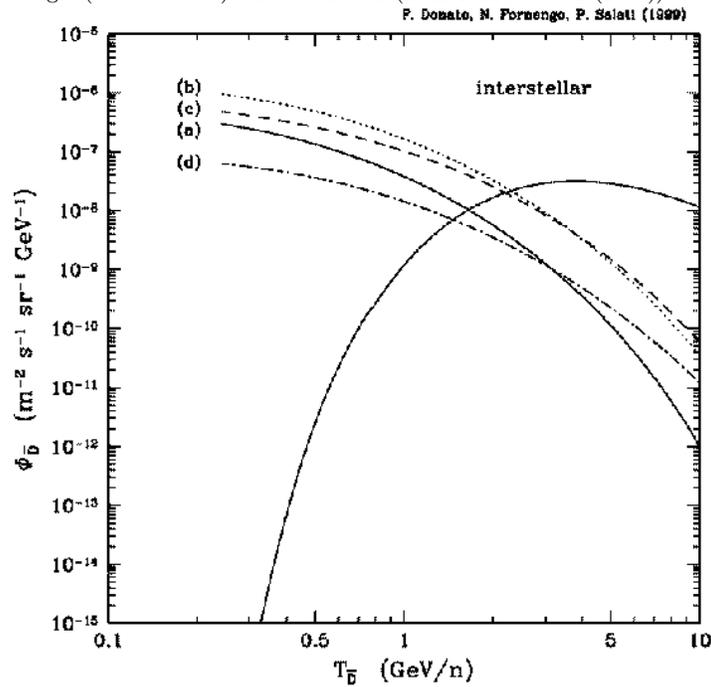}
\end{figure}

The source of background in a primary antideuteron search is the secondary
antideuterons produced in cosmic-ray interactions. The situation is
reminiscent of the primary antiproton searches, where the background is due to
secondary and tertiary antiprotons \citep{simon98, bergstrom99}. However the situation is much
better for a primary antideuteron search. The secondary antideuterons cut off
at much higher kinetic energy than in the case of secondary and tertiary
antiprotons (figure \ref{dbar_flux}). If primary antideuterons are searched for at low
enough energies the probability of contamination by secondary antideuterons
can be made negligible, in contradistinction to the primary antiproton case. 

The antideuteron search provides a nice complement to the direct detection
experiments seeking the neutralino. With their very different sources of
background they can together provide a convincing case for the neutralino
detection. In addition there are WIMP models in which the antideuteron signal
would be present but a measurable signal in a direct detection experiment
would not be obtained. 


\subsection{Spectroscopy of ultralow energy antiprotons}
\label{intro_pbar}

The antiproton spectrum has been the subject of numerous theoretical and
observational papers. This is a vast subject and we restrict our comments to
those areas where GAPS represents a significant improvement over current
techniques and where the discovery space is substantial. An example is the use
of GAPS on an interstellar probe to characterize the ultralow energy
antiproton spectrum. At comparable mass, volume and power consumption GAPS has
almost an order of magnitude more grasp ($A\Omega$ product) than alternatives,
can probe to lower antiproton energies and can do so with superior
discriminatory power against false detections (\S\ref{app_pbar}).

\begin{figure}
\epsscale{0.5}
\caption{The cosmic-ray induced interstellar antiproton flux.\label{pbar_flux}}
\plotone{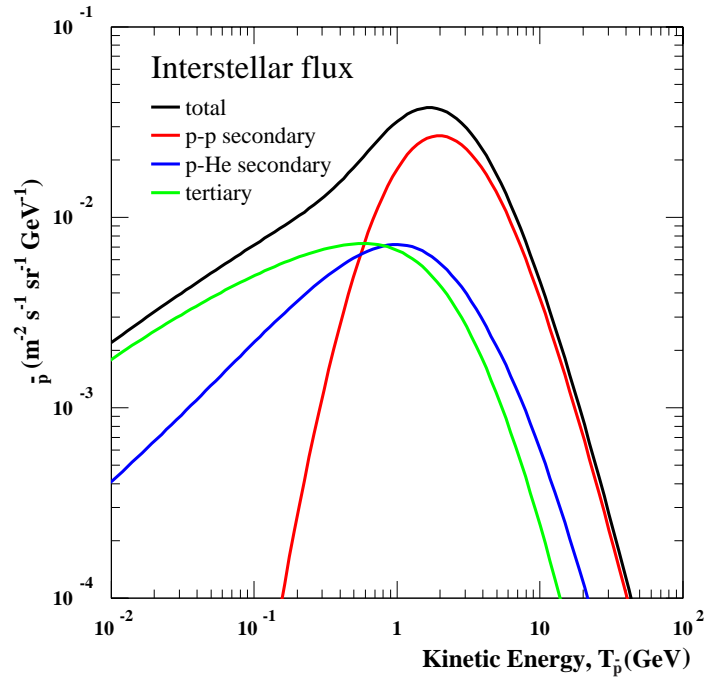}
\end{figure}

The secondary antiprotons are produced through $p+p\rightarrow \bar{p}+X$
reactions and the kinematic suppression of the secondary antiprotons due to
the requirement of 3 protons in the final state, combined with the decreasing
primary proton spectrum, provides a  very characteristic shape for the
secondary antiproton spectrum. This has been probed in many balloon flights
and will be measured by AMS on ISS. The very sharp suppression of the
secondary antiproton flux at low energies provides an opportunity to test
subdominant contributions to the antiproton flux \citep{simon98, bergstrom99, donato01}. These processes
include collisions of primary protons with heavier nuclei (mainly helium) and
energy losses of the secondary antiprotons during propagation (producing the
tertiary antiprotons component). These separate components are shown in figure
\ref{pbar_flux}. The effect of these subdominant components is marked below $\sim$ 150
MeV. But the ability to probe these components is limited by the effects of
solar modulation and the geomagnetic rigidity cutoff inherent in low earth
orbit missions. The modulation prevents antiprotons with energies of less than order
the solar potential ($\sim$ 500 MeV) from reaching the vicinity of the
earth. In order to surmount this difficulty it has been proposed to send a
probe out of the heliosphere \citep{wells99}. In \S\ref{app_pbar}, we show how
a 1-year observation with GAPS on such a probe can detect secondary and tertiary
antiprotons down to 40 MeV. 


Several sources of primary antiprotons could be even larger than the secondary
and tertiary antiproton signal. Antiprotons are produced in the
neutralino annihilation and dominate over the secondary and tertiary
components below $\sim$ 100 MeV for some SUSY models \citep{jungman94,
bottino98}. The evaporation of primordial black holes can produce a signal below
100 MeV which is much larger than either the neutralino induced antiproton
signal or the secondary or tertiary signals, yet can still evade detection on
either balloon experiments or AMS, which operates at much higher energies \citep{macgibbon91, maki96}. 

\subsection{Antihelium}
\label{intro_hebar}

The discovery of a single antihelium atom is compelling evidence for the
existence of an antimatter domain in the universe. Such searches are highly
problematic and thus difficult to motivate. In particular observational
constraints require such domains to be large enough that the antihelium must
travel over great distances through the intergalactic magnetic field and
penetrate into our own galaxy against the galactic wind. There is substantial
uncertainty in the losses that would occur due to these effects
\citep{streitmatter96}. Moreover inflationary cosmologies may provide
additional theoretical biases against the promise of such searches. Recent
attempts to suggest the possibility of a local source of antihelium which has
evaded previous observational limits seem strained \citep{belotsky00}. Nevertheless the
GAPS approach, when implemented as a means to search for the antideuterons,
can provide a 2 order of magnitude improvement in sensitivity over AMS in
setting bounds on the $\bar{He}/He$ ratio. 

\section{The GAPS concept}
\label{concept}

Antimatter spectrometers must identify particle type and energy. Detection
schemes for all proposed and forthcoming missions can be classified in two
categories : magnetic spectrometers and calorimeters. Magnetic spectrometers
measure particle rigidity from which momentum and charge-to-mass ratio can be
determined. Calorimeters identify antiparticles by searching for events with a
total energy deposit equal to twice the rest energy of the antiparticle, as is
obtained from an antiparticle slowed down in the calorimeter and
captured/annihilated by a nucleus. In both schemes the velocity is measured by
the time-of-flight (TOF) method, which is also used to reconstruct the incident
energy. 

Magnetic spectrometers  have an increasing  error in rigidity/momentum
identification below 200 MeV/n due to multiple scattering, which
results in an inaccurate determination of  deflection angle. Also the
effective solid angle for particle acceptance tends to be small for a given
spectrometer surface area since an incident particle has to enter the magnetic
field in a certain angular range for efficient deflection. Calorimeters also
have drawbacks as antiparticle detectors. For instance, in identifying
antiprotons the calorimeter searches for events whose energy deposit is twice
the proton rest energy. But normal hadronic interactions in the calorimeter
initiated by protons, the most common particle incident on the calorimeter,
can produce an energy deposit identical to the antiproton. Therefore the TOF
must be used to exclude those incoming protons whose total kinetic energy is
$2m_pc^2$. The requirement to use the TOF to reject particle velocities
corresponding to this energy, and in the face of enormous proton fluxes, is
challenging. Indeed it requires far better TOF discrimination than has
previously been obtained. Moreover the calorimeter has the disadvantage that
the antiproton signature is not unique - it is simply indicative of a certain
energy deposit. Both the magnetic spectrometer and the calorimeter tend to be
heavy for a given grasp, in the former case because of the magnets and the
latter case because of the need to have a thick enough crystal to completely
contain the antiparticle annihilation gamma-rays. 

\begin{figure}
\epsscale{0.7}
\caption{The operating principal of the GAPS detector using antiprotons as an
example.\label{gaps_view}}
\plotone{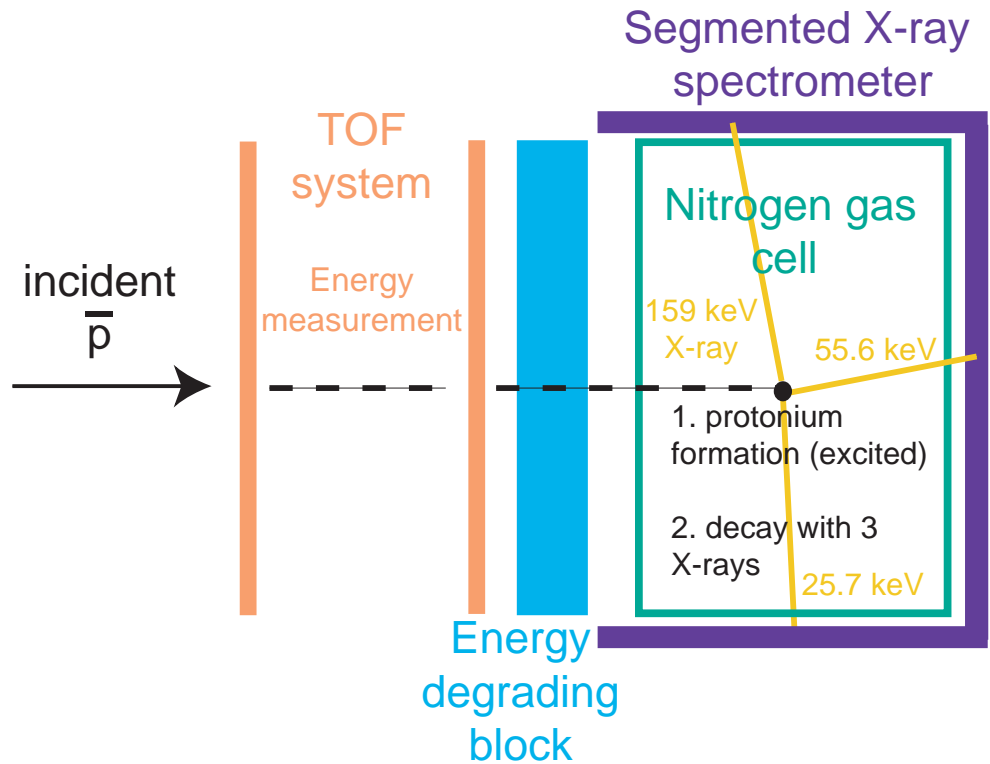}
\end{figure}

A GAPS detector configuration consisting of a single channel (for illustrative
purposes) is shown in figure \ref{gaps_view}. An antiparticle that passes through a TOF
system (which measures energy) is slowed down by $dE/dx$ loss in a degrader
block. The  thickness of this  block is tuned  to select the  sensitive energy
range  of the  detector. The  antiparticle is stopped  in  the gas
chamber, forming an  exotic atom with probability of  order unity. The exotic
atom is in a high excitation state, which deexcites through a process
involving both autoionizing transitions and radiation-producing
transitions. Through proper selection of the target gas and its pressure the
absorption of the antiparticle can be tailored to produce 3-4 well-defined
X-ray transitions in the exotic atom decay chain. Promptly after the release of
these X-rays the antiparticle annihilates in the nucleus producing a shower of
pions. 


The X-rays have energies in the 25-250 keV range so that the gas and the
surrounding gas chamber support structure are optically thin to them. These
X-rays are absorbed in a CZT or NaI spectrometer which surrounds the gas
cell. The coincident signals between the TOF system, the characteristic decay
X-rays and the energy deposition of the pions provide a clean, positive
confirmation of the detection of an antiparticle. The energy of the X-rays
uniquely define the antiparticle mass. 

The above design is only illustrative. A realistic design (\S\ref{application}) contains
$>$ 10 gas cells each surrounded by a segmented X-ray detector and the entire
structure surrounded by the TOF. This configuration provides substantial
stopping power so that antiparticles can be stopped over a broad energy band
through the narrow band stopping power of each cell. Moreover this
segmented design allows for optically thin gas cells (so that the ladder
X-rays can reach the X-ray detector), but permits very high pressure in the small individual
cells. In this manner GAPS can achieve high X-ray detection efficiency along
with total energy bandwidths $\sim$ 0.5 GeV/n.

This technique of antiparticle identification and background rejection
provides a number of advantages, particularly for the detection of low energy
($E<1$ GeV/n) antiparticles. These include extremely high grasp, low weight
per grasp and extremely low probability of false particle identification. In
addition, there is no inherent lower limit on the detectable antiparticle
energy. Instead it is set by the geomagnetic rigidity and the energy losses
in materials (e.g. the TOF) outside the target gas. 

The GAPS concept depends on the detection of the deexcitation X-rays in gas
targets. The definitive experiments were done in noble gases by
\citet{bacher88}. A detailed understanding of the relevant atomic physics is
necessary to apply these results to the optimization of GAPS. We discuss the
relevant issues in the next section. 


\section{Atomic processes in exotic atoms}
\label{atom}

\subsection{Fate of a captured antiparticle in exotic atoms}

Once an antiparticle \xbar is slowed down to an energy of order the ionization
energy of the atom, the antiparticle is captured into an exotic atom replacing
a bound electron. The cross section for trapping of \xbar is of the
order of a molecular  cross-section \citep{beck93}.  The captured antiparticle
is  in a  highly  excited state,  $n_0=(M^*/m_e)^{1/2}$  (assuming that  \xbar
replaces  a  K-shell electron)  due  to  energy  conservation \citep{hayano94}.  For  instance,
$n_0\sim40$ for  antiproton. $M^*$ is  the reduced mass of  the $\bar{X}$-atom
system.   \xbar decays  via a  radiative transition  or an  Auger  process.  A
radiative transition  emits a  photon from the  exotic atom, while  the Auger
process  ionizes  a  bound electron.   The  Auger  process  sets in  when  the
transition energy of  \xbar exceeds the ionization energy  of bound electrons.
Usually,  the Auger  process is  fast compared  to the  radiative transitions.
Since larger  changes in $\Delta n$,  and $\Delta l=\pm 1$ are  preferred for the
deexcitation  (dipole  selection  rule),  \xbar  will drop  to  the  so-called
circular state in  which the orbital angular momentum  takes its largest value
by  the  transition   $(n,l)\rightarrow(n'=n-1,l'=n-2)$.  Although  the  Auger
process may lead \xbar to other  than the circular states with small branching
ratio \citep{hartmann90},  \xbar will be in  a circular state  at an early stage  in its
decaying process.  Decay of circular  states proceeds by $\Delta  n=-1, \Delta
l=-1$ due to the selection  rules for ladder transitions. A simplified version
of the decay process is  : \\[0.5cm] (1) Capture of an antiparticle into an
initial  bound state  $(n_0,  l_0)$\\ (2)  Decay  of the  antiparticle into  a
circular state $(n,n-1)$\\ (3)  Deexcitation via radiative or Auger ionization
of  L, M-shell  electrons\\ (4)  Complete depletion  of bound  electrons\\ (5)
Radiative decay of the  antiparticle with emission of characteristic X-rays\\\
(6) Annihilation of the antiparticle in the nucleus with a pionic shower

In order  to identify  an incident antiparticle,  we measure photons  from the
radiative ladder transitions after all the bound electrons are ionized. Atomic
calculation  is simplified for  the electron-depleted  exotic atoms,  since it
reduces to a two-body problem of the antiparticle and the nucleus. Our concern
is concentrated mainly on transition energy, lifetime of the exotic atom and
quantum number of complete ionization.

\subsubsection{Ladder transition energy}

The photon energy in ladder transition $n \rightarrow n-1$ is given by,
\begin{equation}
E_{ladder}(n)={\tilde Z}^2 \; (\eta
M^*)\Bigg\{\frac{1}{(n-1)^2}-\frac{1}{n^2}\Bigg\}\mbox{ Ry,}
\end{equation}
where $M^*$  is the reduced mass  of \xbar in  units of proton mass  $m_p$ and
 $\eta=m_p/m_e$. $z$ and  $Z$ are the charge of the  incident particle and the
 absorbing  atom  and  we   defined  $\tilde{Z}=zZ$  for  convenience.  Ladder
 transition energies are listed in table \ref{e_pbar}--\ref{e_hebar}.


\subsubsection{Lifetime of an antiparticle in the exotic atom}

The transition  rate,  $\Gamma_{ladder}(n)$   [sec$^{-1}$],  for  ladder  transition
($n\rightarrow n-1$) is given by,
\begin{eqnarray}
\Gamma_{ladder}(n)&=&\frac{4{E_{ladder}(n)}^3}{3\hbar^4c^3}|\langle
n-1,n-2|\vec{r}|n,n-1\rangle|^2\nonumber\\   &=&5.9\times10^{13}  M^*  {\tilde
Z}^4n^{-5}.
\end{eqnarray}
Therefore, the transition time is :
\begin{equation}
\tau_{ladder}(n)  \sim 1.6\times  10^{-14}{M^*}^{-1}{\tilde  Z}^{-4}n^5 \mbox{
sec.}
\end{equation}
When \xbar  replaces an  electron in shell $N_{sh}$ ($N_{sh}=1,  2, 3$
corresponds to  K, L, M-shell respectively),  its initial bound  state is $n_0
\sim N_{sh}(\eta M^*)^{1/2}$. The lifetime $\tau_{\bar{X}}$ is given by,
\begin{equation}
\tau_{\bar{X}}\sim 10^{-5} N_{sh}^6{M^*}^2 {\tilde Z}^{-4} \mbox{ sec.}
\end{equation}
This  is an upper limit to the lifetime which assumes that  all the  transitions are  via
radiative deexcitation. Delay of the annihilation of antiprotons in helium was
observed  by  \citet{nakamura94},  indicating  the formation  of  antiprotonic
helium atoms with lifetime of order $10^{-6}$ sec.

\subsubsection{Principal quantum number of complete ionization}

When the transition energy of \xbar becomes larger than the K-shell ionization
energy of a  bound electron, the Auger process ionizes  a K-shell
electron. The quantum number $n_K$ where electrons are completely depleted is given
by, 
\begin{equation}
E_{ladder}(n_K) = I_K \mbox{ }\left(= Z^2 \mbox{ Ry}\right).
\end{equation}
In this  case, $n_K$ is  independent of $Z$.  $n_K$ is $\sim 15, 19, 38$  for
\pbar, \dbar and \hebar respectively. \citet{bacher88}  observed
strong  suppression  of  $n=15,16$   ladder  X-rays  in  several  noble  gases
illuminated by antiproton beams. All the transitions from $n<n_K$ are
radiative and ladder photons from lower $n$ (e.g. $n=4-7$ for $\bar{p}$) are observed by X-ray detectors. 

\subsection{Relevant atomic processes and their transition rates}

\begin{figure}
\epsscale{0.7}
\caption{Deexcitation path of a captured antiparticle in an exotic atom.\label{decay_path}}
\plotone{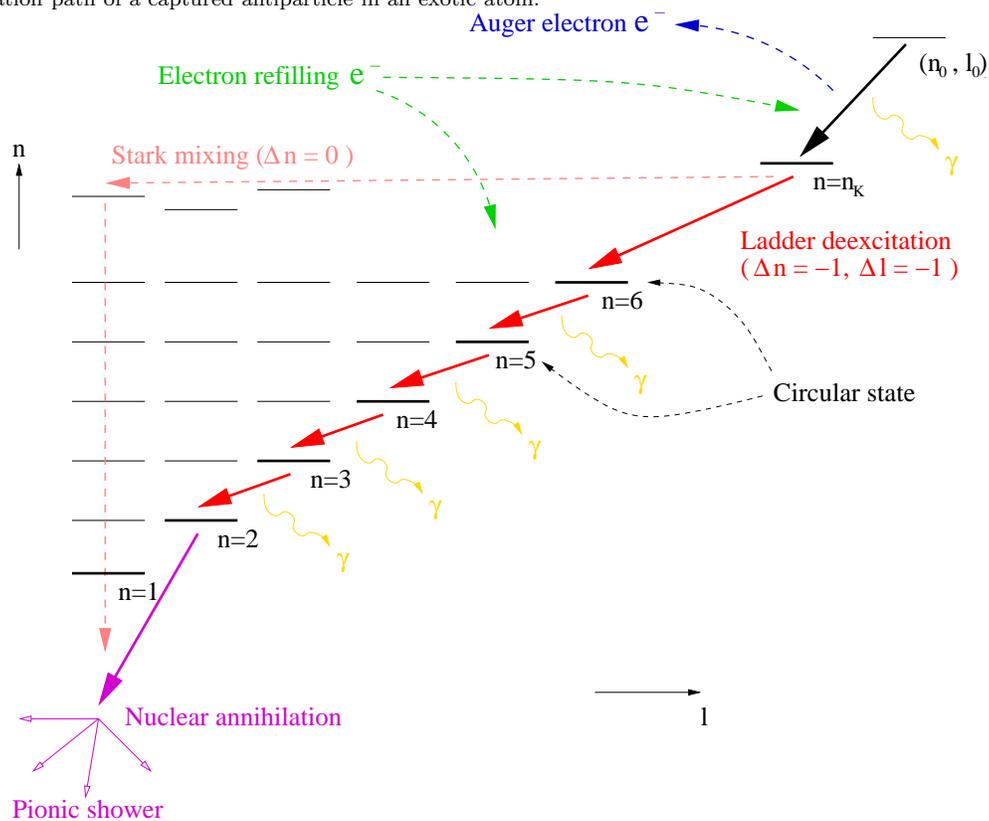}
\end{figure}

\begin{figure}
\epsscale{0.6}
\caption{Competing atomic processes in antiprotonic Argon atom at $\rho=0.5\mbox{ gcm}^{-3}$, $T=0^\circ$C and $\sigma_r=10^{-14}$ cm$^2$. All the
bound electrons are ionized at $n_K=13$ due to the delay of complete
ionization by electron refilling. The rate of Auger ionization was calculated
by Ferrell's formula \citep{ferrell60}. \label{rate}}
\plotone{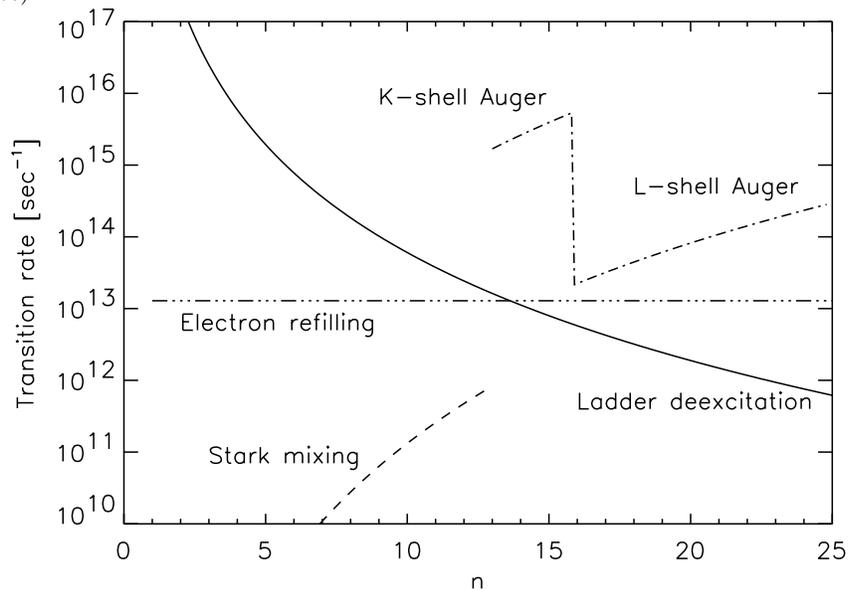}
\end{figure}

In order to increase efficiency for detecting $\bar{X}$, high
density  targets are  desirable for  the  detection medium.  As the  density
increases,  several  atomic  processes  may
interfere  with the identification  of characteristic  X-rays from  the ladder
transitions. Figure \ref{decay_path} schematically shows the deexcitation path
of an exotic atom and figure  \ref{rate} presents the rates of different atomic
transitions discussed in the following section.


\subsubsection{Stark mixing}
A quantum state of \xbar  becomes degenerate after the full ionization of
bound    electrons.  Then  the    electric    field   from    an   adjacent    atom
$\vec{E}=(e/R^3)\vec{R}$,  distorts  different  $l$ quantum  states,  inducing
$\Delta{n}=0$  transitions.  $R$  is  the  intermolecular  distance  or  impact
parameter of  a passing  atom in the  gas. This  transition leads \xbar  to an
$S$-state  followed by  a nuclear  annihilation  or to  an $nS\rightarrow  1S$
radiative transition. The photon  energy from the $nS\rightarrow 1S$ radiative
transition in  high $Z$ materials  is too  high to be  measured by a  thin
X-ray detector. A measure of the transition width caused by the energy shift $\Delta
E_{Stark}  =  \langle n,  n-2  | e\vec{E}\cdot\vec{r}  |  n,  n-1 \rangle$  is
\citep{day59},
\begin{equation}
\omega_{Stark}(R)\sim\frac{\Delta E_{Stark}}{\hbar}=2.2\times10^{13}
{M^*}^{-1} R^{-2} {\tilde Z}^{-1} n^2
\end{equation}
Stark mixing becomes dominant over the  ladder transition when $R \le R_n$, where $R_n$ (in units of Bohr radius $a_0$) is given by,
\begin{equation}
\omega_{Stark}(R_n) \sim \Gamma_{Ladder}(n),
\end{equation} 
or
\begin{equation}
R_n \sim 0.6 {M^*}^{-1} {\tilde Z}^{-5/2} n^{7/2}.
\end{equation}
$R_n$  can  be  larger than  the  intermolecular  distance  in  a solid  or  a
liquid.  This leads to suppression of the ladder X-ray transitions of interest. However, molecules  in a  gas follow  the Boltzmann  distribution and
present different impact  parameters to the exotic atom.  For this collisional
process, the rate of Stark mixing is given by,
\begin{equation}
\Gamma_{Stark}=N_{a}\;(\pi{R_n}^2) \; v.
\end{equation}
$N_{a}$ is the number density of atoms. We represent the velocity of atoms $v$
in the  gas by their thermal  velocity $v_{th}=\sqrt{\case{3kT}{Am_p}}$, where
$A$ is the atomic weight of the gas. Therefore,
\begin{equation}
\Gamma_{Stark}(n)=3.1\times  10^{12} {M^*}^{-2}  {\tilde Z}^{-5}  A^{-3/2} n^7
\,\rho \, {T}^{1/2},
\end{equation}
where $\rho$ and  $T$ are the density $[\mbox{gcm}^{-3}]$  and temperature [K]
of the gas.

\subsubsection{Electron refilling}
When  an exotic atom  is highly  ionized, charge  transfer from  other neutral
atoms  can refill  shells in  the  exotic atom.  Charge transfer  may cause  a
continual  cycle of  Auger ionization  at rates  high enough  to  suppress the
radiative transitions.  The cross-section for charge transfer  from adjacent atoms
is    in    the     range    of    $\sigma_r=10^{-14}-10^{-15}    \mbox{cm}^2$
\citep{ryufuku80}.  When  $n_e$  is   the  number  density  of  electrons  and
$\sigma_{r,-14}$  is the  cross  section  of electron  refilling  in units  of
$10^{-14} \mbox{cm}^2$, the rate of electron refilling is given by,
\begin{equation}
\Gamma_{refill}=n_e  \,\sigma_r \,v_{th}  =  4.7\times10^{13} A^{-1/2}  \rho 
\,{T}^{1/2} \sigma_{r,-14}.
\end{equation}
Metals are  not appropriate  target materials because their  fast Fermi
velocities  ($\sim 10^8$ cm/s) lead to rapid refilling rates.


\subsubsection{Nuclear absorption}

The  strong   nuclear  decay   rate  of  the   $2P$  state  in   protonium  is
$\Gamma_{2P}=35$     meV      \citep{reifenrother89},     corresponding     to
$\Gamma_{2P}=5.3\times10^{13}$ sec$^{-1}$.  Nuclear annihilation for  the $2P$
state  can be  as fast  as its  radiative deexcitation.   The $2\rightarrow1$
ladder transition is  also excluded from observable line  candidates since its
transition energy  is much higher than  the energy band of  a moderately thick
X-ray   detector.   The   strong   interaction   in   D-states   is
negligible \citep{reifenrother89}.

\subsection{Yield of ladder transitions}

The   initial  capture  angular   momentum  $l_0$   is  not   well  understood
presently. However, we can assume a statistical distribution for $l_0$, i.e. a
probability for  the capture of an  antiparticle into states $l_0$  is given by
$\case{2l_0+1}{{n_0}^2}$.  An  antiproton   captured  into  $l_0=0-5$  is  not
relevant, since it  decays into a low  $n$ circular state which  does not
give   several   ladder   transition  photons   observable   by  a thin
X-ray detector. However, the probability of capture into such a low
$l_0$ state is less than 10\%. Coulomb deexcitation, a process 
by which transition energy is transfered  to kinetic energy of the exotic atom
and a nearby  atom, reduces the yield of ladder  transition photons \citep{aschenauer95}. However, the  significance of the Coulomb deexcitation  is still not
well understood and is likely a small effect.

In experiments it has  been shown  that the yield  of ladder  transitions is
dependent  on the  gas pressure.  The yield  of ladder  transitions ($n\ge7$)
was measured to  be $\sim$ 50\% for  relatively low pressure  ($\sim 10^{-2}$
atm) gases forming antiprotonic atoms \citep{bacher88}.  On the other hand, Lyman lines
were  measured  at  high  gas  pressures  ($\le$ 170  atm)  for  muonic  atoms
\citep{jg88}. Hereafter,  we set the  yield of the overall  ladder transitions
$Y_{n_1}=$  50\%  ($n_1$ is the 1st  circular  state  of  the  ladder  transitions  of interest) and  $Y_n=$ 100\% ($n<n_1$, lower $n$ ladder  transitions). We
will investigate the pressure dependence of the yield of ladder transitions at
antiproton beaming facilities.

\subsection{Optimal gas element and density}
\label{density}

\begin{figure}
\epsscale{1.0}
\caption{Maximum gas density $\rho_{max}$ [$\mbox{gcm}^{-3}$] for antiprotons  
(left) and antihelium (right). The temperature is $T = 0 ^\circ\mbox{C}$ for
illustrative purposes.\label{max_density}}
\plottwo{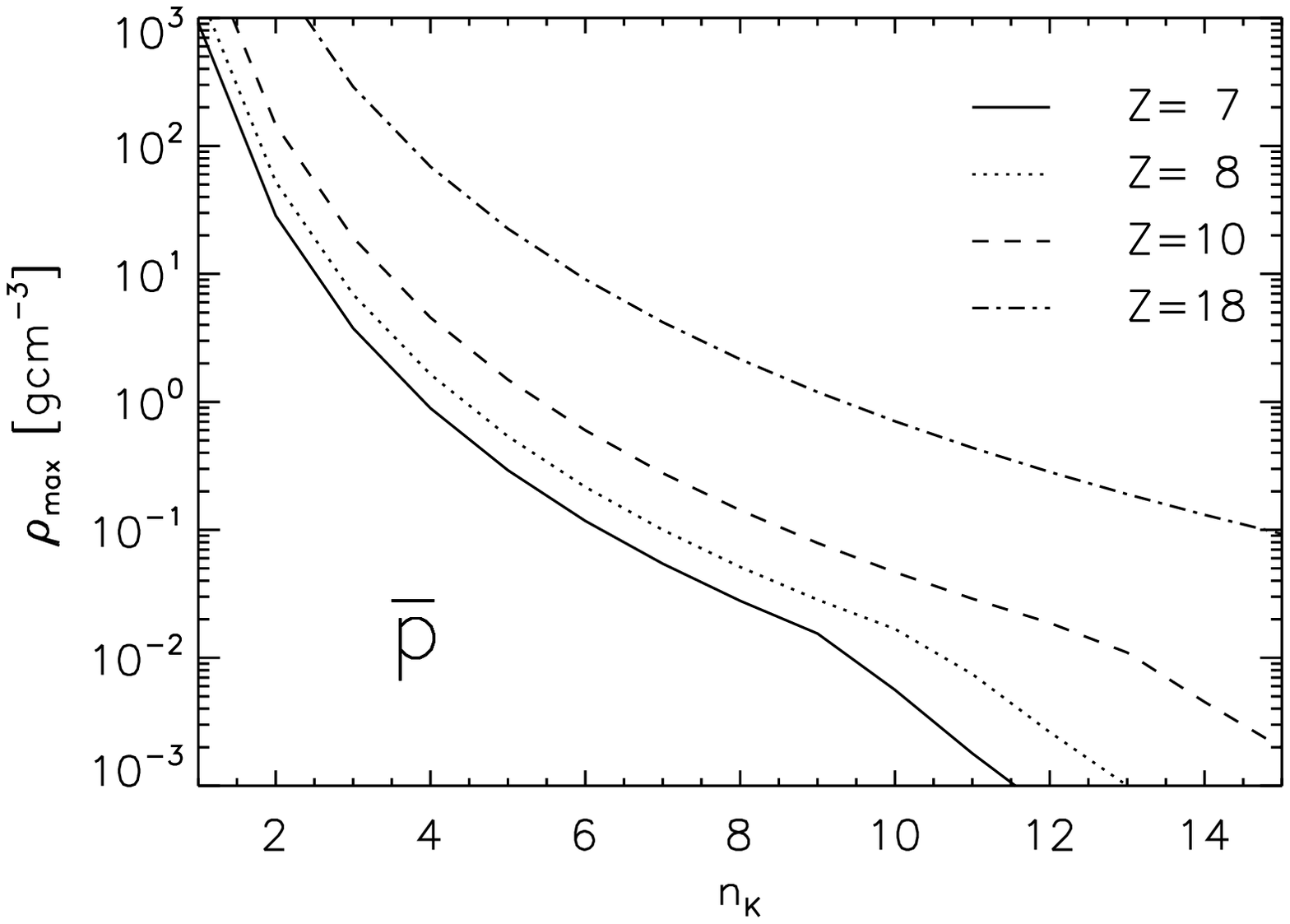}{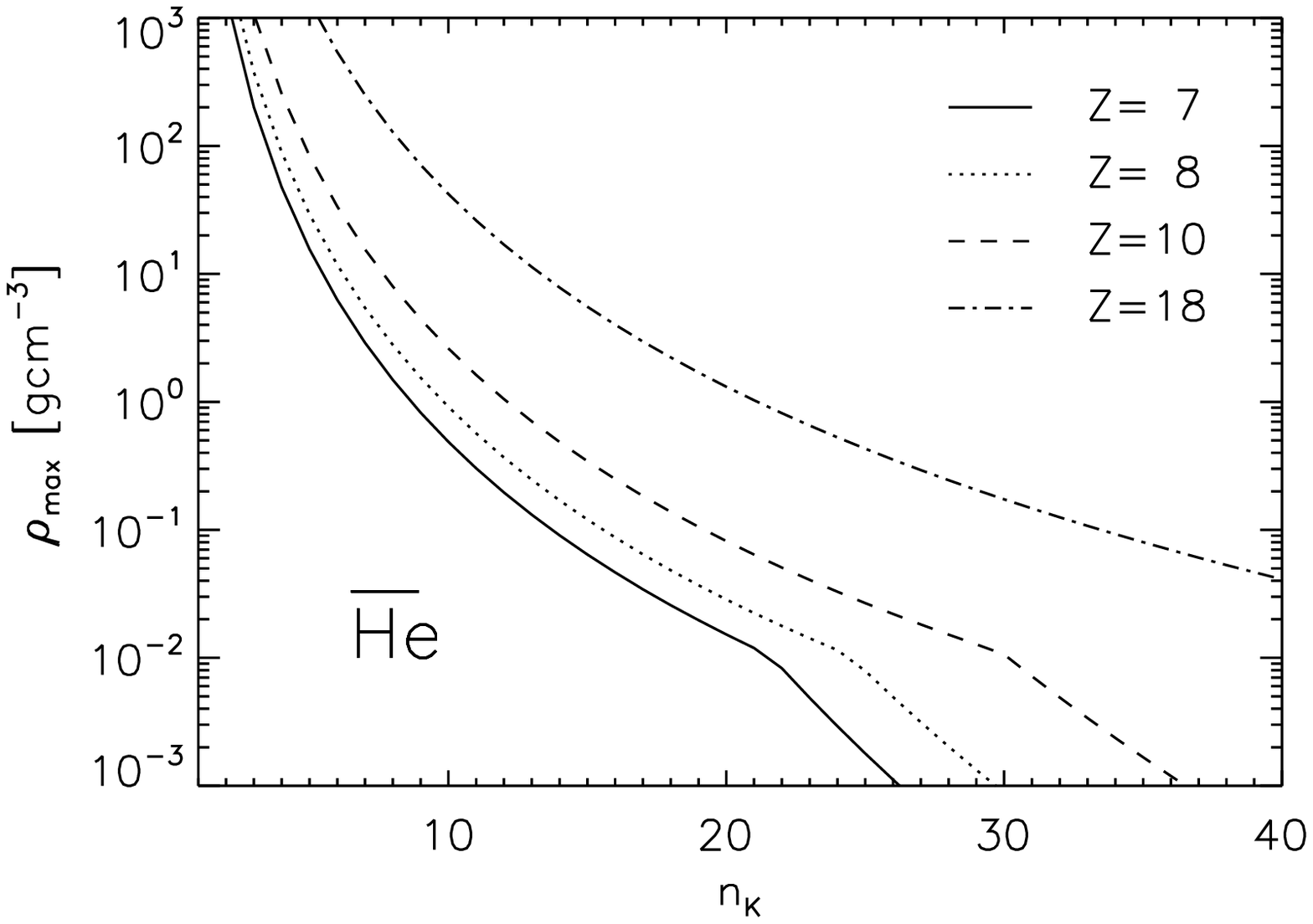}
\end{figure}

Atomic calculation sets an upper  limit on the acceptable gas density required
to generate the characteristic  X-rays from the ladder transitions. At
high density, Stark  mixing and electron refilling can  suppress the radiative
deexcitations. However, Stark  mixing does not  set in
when  electrons are present  in the  exotic atom and so  the electron  refilling can
suppress the Stark mixing as well. Therefore, the optimal situation is for the
ladder transition rate to be faster than  Stark mixing rate by the time electrons no
longer refill the  exotic atom (more precisely, refill  the K-shell). Refilling
of  the L-shell  can  still be faster  than ladder  transitions. However,  at
$n\le7$, the  fluorescence transitions of  L-shell electrons to the K-shell become
slower than  the ladder  transition in antiprotonic  atoms. This  condition is
explicitly described as,
\begin{equation}
\Gamma_{Ladder}(n_K) > max\,\{\Gamma_{Stark}, \,\Gamma_{refill}\}.
\end{equation}
Therefore,
\begin{eqnarray}
\rho\mbox{ [gcm}^{-3}]  &<& min\,  \bigg\{1.9\times 10^1 {M^*}^3  {\tilde Z}^9
A^{3/2} {n_K}^{-12}  T^{-1/2}, \nonumber\\ &  &1.3 \, M^*{\tilde  Z}^4 A^{1/2}
{n_K}^{-5} T^{-1/2} {\sigma^{-1}_{r,-14}}\bigg\}.
\end{eqnarray}
We plot the maximum density for several gases in figure \ref{max_density}. Hydrogen and helium
are ruled out due  to their fast Stark mixing rate at  the gas density desirable
for stopping  antiparticles. On the other hand,  Kr and Xe are  not optimal 
because  of  high  photo-absorption  of  ladder X-rays  observable  with
X-ray detectors. Therefore,  we limit our choice of  the gas to N$_2$,  O$_2$, Ne and
Ar. Gas selection will be discussed further in the following sections.



\section{Detection efficiency}
\label{sec_qe}

We  assume  a  cubic  detector   for  a  preliminary  calculation  of
detection efficiency. A detailed description of the  detector design will be found in \S\ref{design}. Hereafter, we denote the kinetic energy of an incident particle as
$E_{\bar{X}}$,  its  incident direction  as  $\hat{n}=(\theta,  \phi)$ and  an
incident point  on the  detector surface $S$  as $\vec{r}=(x,y,z)$.  The total
detection efficiency is the product  of the efficiency of forming an exotic
atom   in   the   gas   ($\epsilon_{cap}$)   and   detecting   ladder   X-rays
($\epsilon_{\gamma}$).
\begin{equation}
\epsilon_{tot}(E_{\bar{X}},\hat{n},
\vec{r})=\epsilon_{cap}\cdot\epsilon_{\gamma}.
\end{equation}
$\epsilon_{\gamma}$ is  decoupled from $\epsilon_{cap}$ since  it depends only
on the photon  energies and the point $C$ at which  \xbar is captured. Effective
grasp $A\Omega(E_{\bar{X}})$ [m$^2$sr] is given by,
\begin{equation}
A\Omega(E_{\bar{X}})=\int{d\Omega}\int{dS}\;\epsilon_{tot}(E_{\bar{X}},
\hat{n}, \vec{r}).
\end{equation}
We define  an energy band  by integrating $A\Omega(E_{\bar{X}})$ over energy
and dividing by  the peak
$A\Omega(E_{\bar{X}})$.

\subsection{Efficiency of forming an exotic atom}

The quantum efficiency of forming an exotic atom in the gas is given as,
\begin{equation}
\epsilon_{cap}=\epsilon_{cap}(E_{\bar{X}},  \hat{n},  \vec{r}) =  P_{cap}\cdot
(1-P_{ann})\cdot \epsilon_{rig}.
\end{equation}
A  capture point  $C$ is  uniquely determined  for a  given set  of parameters
$(E_{\bar{X}}, \hat{n},  \vec{r})$. $P_{cap}$ is  defined as the probability of
capturing  antiparticles in  the gas  obtained  by subtracting  the capture
fraction in the  X-ray detector. Since the cross-section of  the antiparticle capture
is rather insensitive  to $Z$ \citep{cohen00}, we simply assume  $P_{cap}$ is the
ratio of   the  gas   column   density  to   the   total  column   density  in
the detector.  $P_{ann}$ is  a probability  of  the direct  annihilation with
the nucleus (\S \ref{annihilation}). $\epsilon_{rig}$ represents the effect of
the geomagnetic rigidity cutoff (\S \ref{rigidity}).

\subsubsection{Stopping power and range of antiparticles}

We  adopt data  for stopping  power  and range  from the  NIST PSTAR  database
\citep{pstar}. Data from PSTAR are  applicable to antiprotons in the so-called
Bethe-Bloch  regime ($E_{p}  >$  10 MeV).  For \dbar and \hebar,  we used  a
well-known scaling law \citep{leo90}.

At low  energy ($E_{\bar{p}}<$ 10 MeV),  there are several  minor effects that
can cause ranges  for antiprotons  to deviate from the proton  data. The difference  in the
sign of  charge affects the stopping  power below 1  MeV \citep{barkas63}. The
stopping power  for antiprotons becomes roughly  twice as large  as protons at
the  Bloch peak  ($\sim$ 1  MeV) \citep{adamo93}.  The estimated  deviation in
range is about  1\% for $E_{\bar{p}}=$ 10 MeV.   Multiple-scattering by atomic
electrons  modifies  the  direction  of  incident  particles  (significant  at
$E_{\bar{p}}<$ 1 MeV). Hence, effective range will be shorter than the mean range
assuming a  straight line  projectile. Due  to its heavy  mass compared  to an
electron,  the estimated  error is  less than  1\%  for a  10  MeV antiproton
\citep{pstar}.  The   statistical  distribution  in   range,  so-called  range
straggling,  must be  taken  into account.  A  parameter $\kappa=\bar\Delta  /
W_{max}$, where $\bar\Delta$ : the mean energy loss and $W_{max}$ : the maximum energy
transfer in a single collision, is much larger than 1 for the typical detector
configuration. This implies  that the range straggling is  well described by a
Gaussian distribution in the thick absorber approximation \citep{leo90}.
\begin{equation}
f(x,\Delta) = \exp \left ( \frac{-(\Delta-\bar\Delta)^2}{2\sigma^2}\right),
\end{equation}
where
\begin{equation}
\sigma^2 =0.1569 \; \frac{Z}{A} \;\rho\; x \;\mbox{[MeV}^2\mbox{]}.
\end{equation}
$x$  is the  distance over  which  a particle  propagates in  a material  with
density  $\rho$.  For  a material  with  thickness  1  gcm$^{-2}$ and  50  MeV
antiprotons, the estimated error is less than 1\%.

\subsubsection{Probability of the direct \xbar annihilation with the nucleus}
\label{annihilation}

Cross  sections for  direct  annihilation of  antiprotons  have recently  been
studied. The  lowest energy experiment  was done at $E_{\bar{p}}=1-20$  MeV by
the   OBELIX    collaboration   \citep{obelix96}.   Another    experiment   by
\citet{bruckner90} provided data at $E_{\bar{p}}=20-200$ MeV. Although a $1/v$
law is valid at high energies,  this is modified by Coulomb attraction between
the  antiparticle  and  the nucleus  at  low  energy.  A  general form  for  the
annihilation cross section is \citep{kurki-suonio00},
\begin{equation}
\sigma_{ann}=\sigma_0 \cdot C(v^*, \tilde{Z})/v^*,
\end{equation}
where
\begin{equation}
C(v^*,  \tilde{Z})=\frac{2\pi \tilde{Z} {\alpha  c}/v^*}{1-exp(-2\pi \tilde{Z}
{\alpha c}/v^*)}.
\end{equation}
$v^*$ is the velocity of an incident  particle in the center of mass frame. We
fit  the above  formula to  the experimental  data,  obtaining $\sigma_0=12.7$
mb. The probability of direct annihilation when $\bar{X}$ is decelerated
from $E_{\bar{X}}=E_0$ to $E_1$ is given by,
\begin{eqnarray}
P_{ann}&=&1-e^{-\tau_{ann}},\\      \tau_{ann}&=&     \int_{E_0}^{E_1}     N_a
\;\sigma_{ann}(E) \;\left(\frac{dE}{dx}\right)^{-1} \; dE.
\end{eqnarray}
Since  the  stopping  power  is  almost  independent  of  the  material  type,
$\tau_{ann}$ is insensitive to the atomic number $Z$. Loss of antiparticles by
direct annihilation is $5-10$\% in the typical configuration of GAPS.

\subsubsection{Geomagnetic rigidity}
\label{rigidity}
The efficiency of particle detection depends on the orientation of the
detector and particles with respect to the geomagnetic field. The minimum rigidity  at some geomagnetic
latitude $\varrho$ and geocentric radius $R$ is given by,
\begin{equation}
R_{min}=\frac{\mu_{\oplus}}{R^2}\frac{\cos^4{\varrho}}{[(1+\cos{\theta}\cos^3{\varrho})^{1/2}+1]^2},
\end{equation}
where     $\mu_{\oplus}$    is     the    Earth's     dipole     moment    and
$\mu_{\oplus}/R_{\oplus}^2  =  60$  GV \citep{zombeck82,donato00}.  $\theta$  is the  angle  between  the
direction  of  arrival of  the  particle  and the  tangent  to  the circle  of
latitude.  For  a  given  orbit,  we  computed  $\epsilon_{rig}  (E_{\bar{X}},
\Omega_{det})$ as  a fraction of the  observation time in which  a particle of
kinetic  energy less  than $E_{\bar{X}}$  can  reach the  detector within  its
viewing  angle $\Omega_{det}$. For  instance, assuming  the detector  sees the
entire 
sky,  $\epsilon_{rig}(E_{\bar{D}}=1\mbox{ GeV/n})  = 0.2$  on the  ISS orbit
($52^\circ$  N), while  it is  increased  to 0.4  at $70^\circ$  N where a high
latitude space mission is possible.

\subsection{Quantum efficiency of photon detection}

The  quantum   efficiency  for   detecting  exotic atom photons  $\gamma_i$   with  energy
$E_{\gamma_i}$ at a point $C$ is given by,
\begin{eqnarray}
\epsilon_{\gamma}(C)&=&\prod_{i}\epsilon_{\gamma_i}(E_{\gamma_i}, C),\\
\epsilon_{\gamma_i}(E_{\gamma_i}, C) &=&  Y_{n_i}  \cdot e^{-\tau_{\gamma_i}}
\cdot \epsilon_{det}(E_{\gamma_i}).
\end{eqnarray}
$Y_{n_i}$   :  Yield   of   ladder  transition   $n_i  \rightarrow   n_i-1$.\\
$\tau_{\gamma_i}$ :  Optical depth of a  photon $\gamma_i$ in the  gas and the
pressure vessel.\\ $\epsilon_{det}(E_{\gamma_i})$ : Quantum efficiency of photon detector.

We adopt  data for  photo attenuation lengths  [g$^{-1}$cm$^2$] from  the NIST
database  \citep{xcom}. We  simply assume  the attenuation  of photons  in the
photon   detector  is  $\epsilon_{det}(E_{\gamma_i})$.   We take angle-averaged  values   for
$\epsilon_{\gamma_i}$ since  the direction in  which a photon is  emitted from
the exotic atom is random.


\section{Detector design}
\label{design}

\begin{figure}
\epsscale{0.8}
\caption{Schematic view of the detector consisting of numerous cubic
cells. Each cell is a gas chamber surrounded by a pressure vessel and X-ray detectors.\label{side_view}}
\plotone{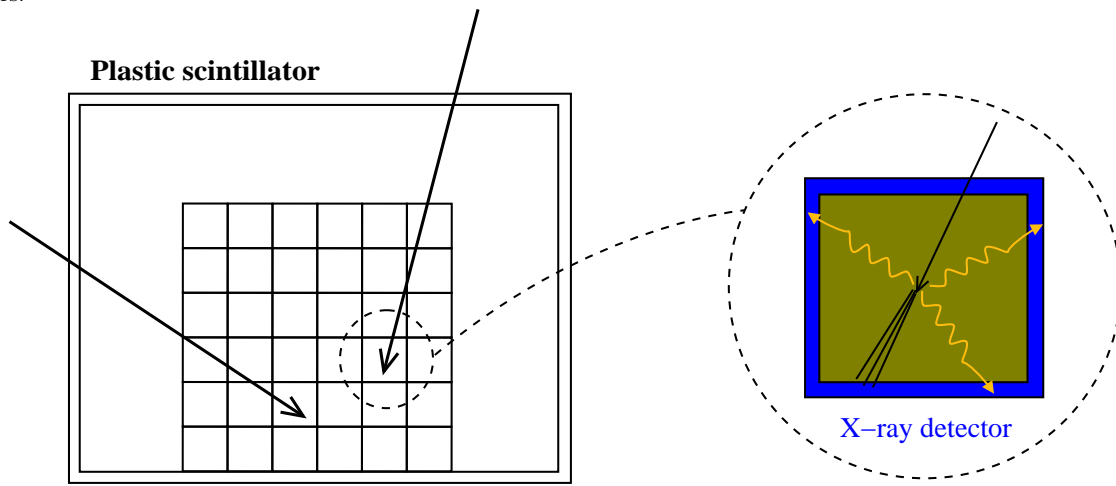}
\end{figure}

The effective grasp  and the band width depend mainly (or  only) on the target
gas column
density. High column density provides a larger acceptance. On the other hand,
the  gas column  density  is limited  by  the photo-attenuation.  In order  to
overcome  this intrinsic  difficulty, we  have  designed a  cubic detector  of
dimension  $L$ [m]  segmented into  numerous  cells. A  cell of  size $l$  [m]
consists of gas  surrounded by a X-ray detector  (figure
\ref{side_view}). While the  overall gas column  density is large, photons  emitted from an  exotic atom
travel only short distance to reach the X-ray detectors surrounding the cell.


A gas column density of 1 gcm$^{-2}$ is selected so that the observable X-rays
do not undergo serious photo-absorption or Compton scattering in the gas. We need
to sustain a gas  pressure as high as $\sim50$ atm in the gas chamber. By  use
of low $Z$ high strength material (e.g. carbon fiber, reinforced plastic etc.), photo-attenuation in
the pressure vessel  wall is  negligible. On the other
hand,   optimization  of   the  thickness   of   the  X-ray detectors  is   more
complicated. The X-ray detector must  be thick enough  to absorb the
ladder X-rays, but thin enough to result in negligible absorption of
antiparticles. Compromise between  these  two  factors  results  in  the
X-ray detector column  density  about  1
gcm$^{-2}$. Outside  the X-ray detector lattice,  we locate two plastic  scintillators of
the total thickness 1 cm separated  by 25 cm for the velocity measurement with
a time resolution $\sim$ 50 $\mu$s.

\subsection{Energy band}
\label{eband}

The  outermost  X-ray detector   and  the  plastic  scintillators  are
sufficiently thin  for 50 MeV/n  antiparticles to penetrate.  With degraders
for slowing down high  energy particles, we can  have at most
5 energy channels corresponding to each surface of the cubic detector except
the bottom. While the column density  of the gas and the X-ray detector of
a cell is fixed  to $\sim 1$ gcm$^{-2}$, the size of the cells  as well as the gas
density  is   determined  for different  configurations  by taking  into
account other constraints  such as the maximum weight (or  size) allowed for a
 given mission. The typical value for the total column
density is $5-10$ gcm$^{-2}$, corresponding  to a band width of $50-70$ MeV/n
in each channel.

\subsection{Background}
\label{background}

Detection of  antiparticles requires extremely reliable  identification in the
presence of  enormous particle backgrounds. The antiproton  flux is $\sim10^5$
times  lower  than the  proton  flux, while  $\sim$  1  antideuteron might  be
observed  for every  $\sim10^9$ protons  and $\sim10^5$  deuterons.  There are
several  kinds   of  ``background''.  For  instance,   a  cosmic-ray  produced
antideuteron can be  mistaken for a neutralino produced  antideuteron. This is
not really a  misidentification of the antiparticle but  rather its production
mechanism.  The  backgrounds   in  this  section  are  those   that  cause  an
antiparticle  to be misidentified  as another  particle or  antiparticle. This
background is  a set of X-rays  whose energies exactly  mimic the antiparticle
ladder X-rays  and which  occur in a  time window  when a candidate  event has
triggered  the TOF. For  instance a  proton may  produce a  TOF signal  in the
energy band of interest while activation of the detector volume by cosmic-rays
leads to the emission of 3  or more X-rays or even $\beta$-particles mimicking
the X-rays  of protonium during  the time the  TOF trigger occurs.  This would
lead  to  the  misidentification  of  the proton  as  an  antiproton.  Similar
considerations apply  to the antideuteron.  The background X-rays can  also be
produced  by hadronic  or pionic  interactions leading  to radiation,  such as
bremstrahlung,  or  even  a  direct  ionization energy  deposit  of  the  same
magnitude as that produced by a ladder X-ray transition.

The proper consideration of background is complicated and the subject of
ongoing study. We only mention some of the key issues here. GAPS should ideally
operate at high latitudes to enhance the flux of antiparticles. But that also
means very high proton fluxes. The $\gamma$-ray and $\beta$-particle
background from spallation and activation by cosmic-rays must be estimated. In
addition the pionic showers produced during annihilation of an antiparticle
are a source of secondary X-rays which can confuse the identification of one
antiparticle for another or serve as a source of background leading to a
misidentification of a particle in an adjacent detection cell. Much more work
is required to understand the sources of background and their exact
impact. However we have attempted to estimate the background for the most
aggressive use of GAPS; an antideuteron experiment in a highly-inclined orbit
(\S\ref{app_dbar}). We used data from the balloon experiment GRATIS \citep{keck01} to scale
to the satellite experiment using procedures well-described in the literature
\citep{harrison01}. 
The typical background for this orbit of $\sim10$ cts/cm$^2$/s in
the relevant energy band of the ladder transitions is almost 2 orders of
magnitude higher than in an experiment at the mid-latitudes due to effects of
increased particle flux on activation, spallation and secondary $\gamma$-ray
background in the payload. This background estimation indeed includes all the
potential sources because it is based on measured X-ray background in previous
space and balloon experiments. We have extrapolated background rates at higher
latitude orbit by estimating increase of cosmic-ray rate based on geomagnetic
rigidity which is found in \S\ref{rigidity}. For the energy resolution of
the CZT and a 10 $\mu$s time resolution the probability of detecting 3 photons in the right
energy and time window to mimic an antiproton for the $\sim$ 1 m$^2$ area of
the design below, is $\sim8\times10^{-9}$. This is sufficient to ensure
negligible misidentification of protons as antiprotons. 

The same calculation can be done for antideuterons. Here, there is a
significant effect since protons, which are the dominant source of
misidentifications, can be rejected with modest efficiency by the TOF because
their energy deposit differs from that of an antideuteron. This is sufficient
to reach the $\sim10^{-12}-10^{-14}$ rate of accidental misidentification per
proton, which is required. Other sources of background leading to
misidentifications are less important than those considered here. We have
probably underestimated the true rejection power. Much of the background we
considered is generated when activation $\gamma$-rays of higher energy than
the ladder transition X-rays Compton-scatter in a detection cell, leading to a
partial energy deposit mimicking a ladder transition X-ray. The scattered
$\gamma$-ray will be absorbed in another cell allowing a possibility of using other detection
cells as part of a veto system for such events. Pionic X-rays are also
produced in the nuclear annihilation and when used in coincidence with ladder
X-rays can provide additional discriminatory capability. We have not yet
investigated their potential. 

We should also note that GAPS has no problem distinguishing antiprotons from
antideuterons because the ladder transition X-rays are uniquely identified in
the two cases. This scheme is dramatically different than the approach of
\citet{wells99}. It does not require any calorimetric signature for 
identification. The fact that the relevant X-ray signature is ``constrained''
in one detection cell provides additional flexibility for rejecting background
that needs to be investigated. 

We primarily envision GAPS being employed with high energy resolution CZT
detectors. However alkali halide scintillator crystals such as NaI(Tl) provide
a simpler and cheaper alternative X-ray detector. The background rejection
capability of NaI(Tl) is comparable or even somewhat better than CZT. The
poorer energy resolution is more than offset by the superior temporal
resolution compared to CZT. If NaI is used the energy resolution is sufficient so that
the antiproton and antideuteron can be resolved cleanly through the comparison
of the highest energy of the 3 ladder X-rays of interest in each case (refer
to table
\ref{e_pbar}, \ref{e_dbar}). The 2
lowest energy X-rays from the antiprotonic and antideuteronic atoms cannot be
resolved from each other in NaI. The preference for CZT is primarily based on
the belief (to be studied) that it will be easier to implement the highly
segmented readout system and multi-cell detection geometry. In addition, the
accuracy to which the 3--4 ladder X-rays can be measured in CZT provides a very
powerful positive confirmation of the presence of antimatter. 


\section{Potential application of the GAPS detector}
\label{application}

We discuss a few potential applications of GAPS based on model calculation of
instrument performance.

\subsection{Antideuteron}
\label{app_dbar}

\begin{figure}
\epsscale{0.7}
\caption{Effective grasp [m$^2$sr] for the detection of the antideuteron. The
detector configurations for high latitude mission are present
in table \ref{dbar_table}. This is for the lowest energy channel. The sharp cutoff at low energy is due to the particle
degradation by the outermost CZT detector and the plastic scintillators.\label{dbar_grasp}}
\plotone{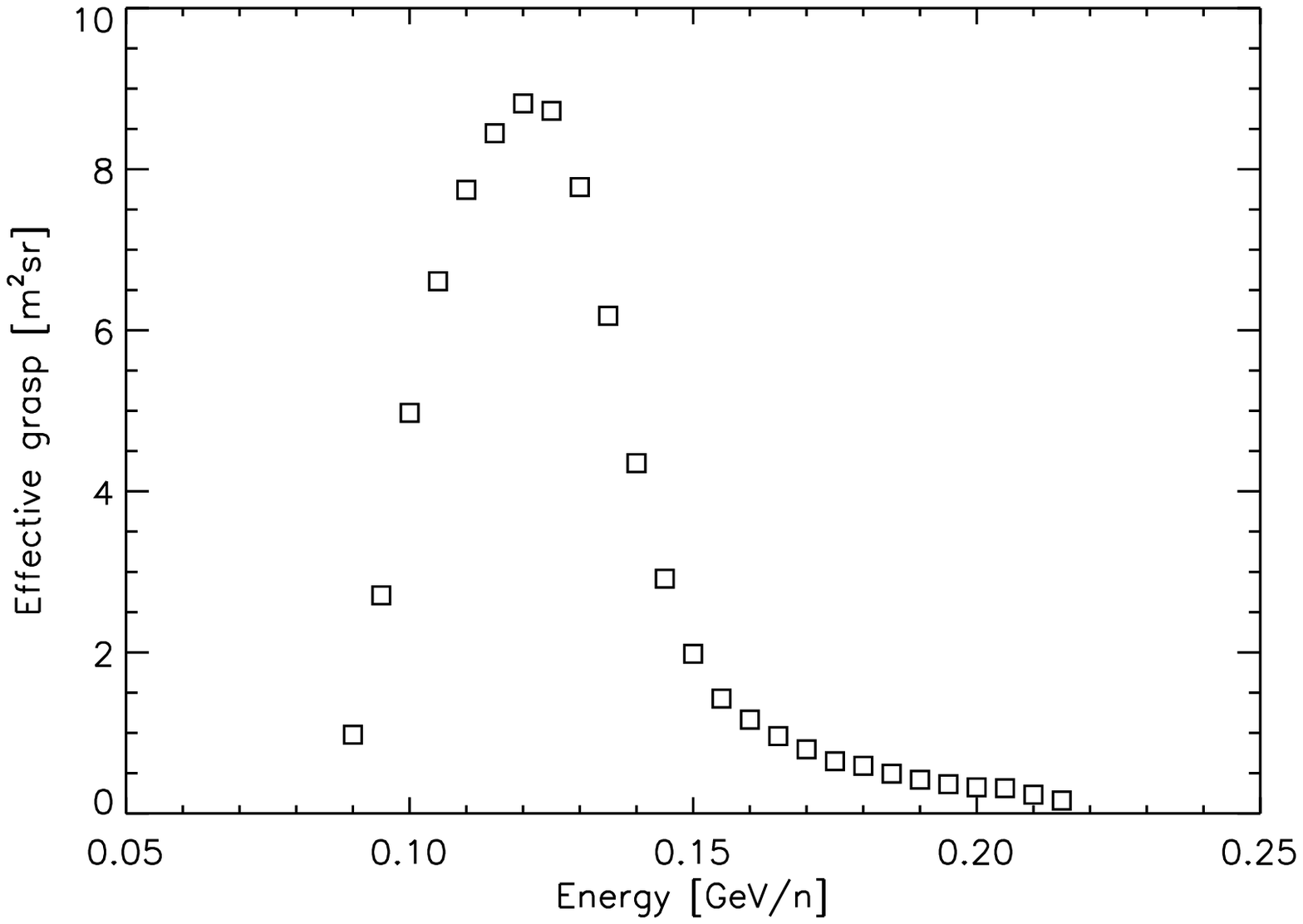}
\end{figure}

As discussed in \S\ref{intro_dbar} a sensitivity $\sim
10^{-9}$m$^{-2}$sr$^{-1}$GeV$^{-1}$sec$^{-1}$ is  required below 1 GeV/n. A
multiyear space mission is required to achieve such sensitivities. We have
simulated a high inclination (70$^\circ$N) mission. The high latitude mission
(HLM) is
advantageous in reducing the geomagnetic rigidity cutoff. Assuming nitrogen
gas and 27 GAPS cells of size $l=160$ cm surrounded by pixellated CZT detectors, we achieve a
peak grasp of 9.0 m$^2$sr for antideuterium over an energy band of 0.1 to 0.4
GeV/n with the detector size of $L=5$ m and a total mass of less than 10,000 lbs (table \ref{dbar_mission}). This results  in   the   sensitivity   $2.6\times10^{-9}$m$^{-2}$sr$^{-1}$GeV$^{-1}$sec$^{-1}$ in 3 years,  20  times  better  than
AMS (table \ref{dbar_table}).  A model calculation of the effective grasp for the proposed experiment
is in figure \ref{dbar_grasp}. In addition to superior sensitivity to AMS, we also note that
the cost of a GAPS instrument with 20 times the sensitivity of AMS is about
an order of magnitude less. The type of mission described here could readily be
executed as a modestly sized NASA ``MIDEX'' class mission. 


\subsection{Antiproton}
\label{app_pbar}

The  recent  BESS  measurements
detected only  a few antiprotons  below 200  MeV, with resultant large error
bars.  To detect a statistically significant number of antiprotons in several
low energy bands at $E_{\bar{p}}<0.5$ GeV requires a sensitivity of $\sim
10^{-3}$ m$^{-2}$sr$^{-1}$GeV$^{-1}$sec$^{-1}$.   We simulated a balloon-borne
experiment performed at high latitude where the rigidity cutoff is $<$ 0.5
GeV. The detector size is $L=2$ m and the overall detector column density is 9
gcm$^{-2}$. It consists of $5\times5\times5=125$ cubic cells of size  $l=40$ cm. The estimated total weight of the
detector is less than 4000 lbs. Neon  gas is  chosen  since 3  ladder  X-rays (29.12,  53.60,  115.8 keV)  are
observable by a 1 gcm$^{-2}$ thick NaI  detector (preferred over CZT on a
balloon experiment because it is cheaper). The  effective grasp is
2.4 m$^2$sr with a bandwidth $\sim 70$ MeV for each energy  channel. This is 4
to   8  times  larger  than  the   geometrical  grasp  of  the  AMS  and  BESS
 experiments respectively.   Acceptance  for   each  energy   channel   is  $1.4\times10^4$
m$^{2}\,$sr$\,$GeV$\,$sec  for  1 day  observation  time.  In several  balloon
flights over the 11 year solar cycle, one could study the effect of solar
modulation  or any excess  from the  predicted secondary  flux at  5 different
energy bands over $E_{\bar{p}}=120-400$ MeV.
	 
The most effective means of probing  exotic sources of antiprotons such as the
evaporation  of  primordial black  holes  \citep{macgibbon91,  maki96} or  the
annihilation of neutralinos \citep{jungman94, bottino98},  is to send an instrument  into deep space beyond
the  heliosphere.  An instrument  in  deep space  is  far  removed from  solar
modulation effects  and geomagnetic rigidity cutoff inherent  in low earth
orbit missions. This permits   investigation of  low  energy  antiprotons,
especially  at $E_{\bar{p}}<100$  MeV.    We  have  designed  a  detector   with  two  energy
channels. A measurement  in the lower energy band $40-60$  MeV, where the flux
due  to p--p  interactions is  negligible, is  to obtain  a clean  signature of
the p+He  and  tertiary antiprotons  or  these in  combination with  a
possible neutralino signature, and/or primordial black hole signature. The second channel at 100 to 120 MeV anchors the
antiproton flux and spectral shape to higher energy observations.

The  detector   is  a  cubic  cell   of  size  $L=6$  cm   comparable  with  a
recently-proposed BGO detector by \citet{wells99}. Argon gas is mandatory from
the  constraint on  the gas  density discussed  in \S\ref{density}.  All six
surfaces of the  cube see the whole  sky and they are divided  into two energy
channels by  properly mounting degraders.  The effective grasp is  65 cm$^2$sr
for  each channel,  6  times larger  than  the geometrical  grasp  of the  BGO
detector while the  estimated weight of the detector is a few kilograms; less
than the BGO detector.  In addition,  the background  rejection power of GAPS
is far better. In  table \ref{interplanetary}, we  estimated the overall  counts from
the  intensities for  interstellar antiprotons.  GAPS is  able to  constrain a
source of antiprotons by 1 year observation.


\subsection{Antihelium}
\label{app_hebar}

\begin{figure}
\epsscale{0.6}
\caption{Comparison of the upper limits of the ratio $\bar{He}/He$.\label{helimit}}
\plotone{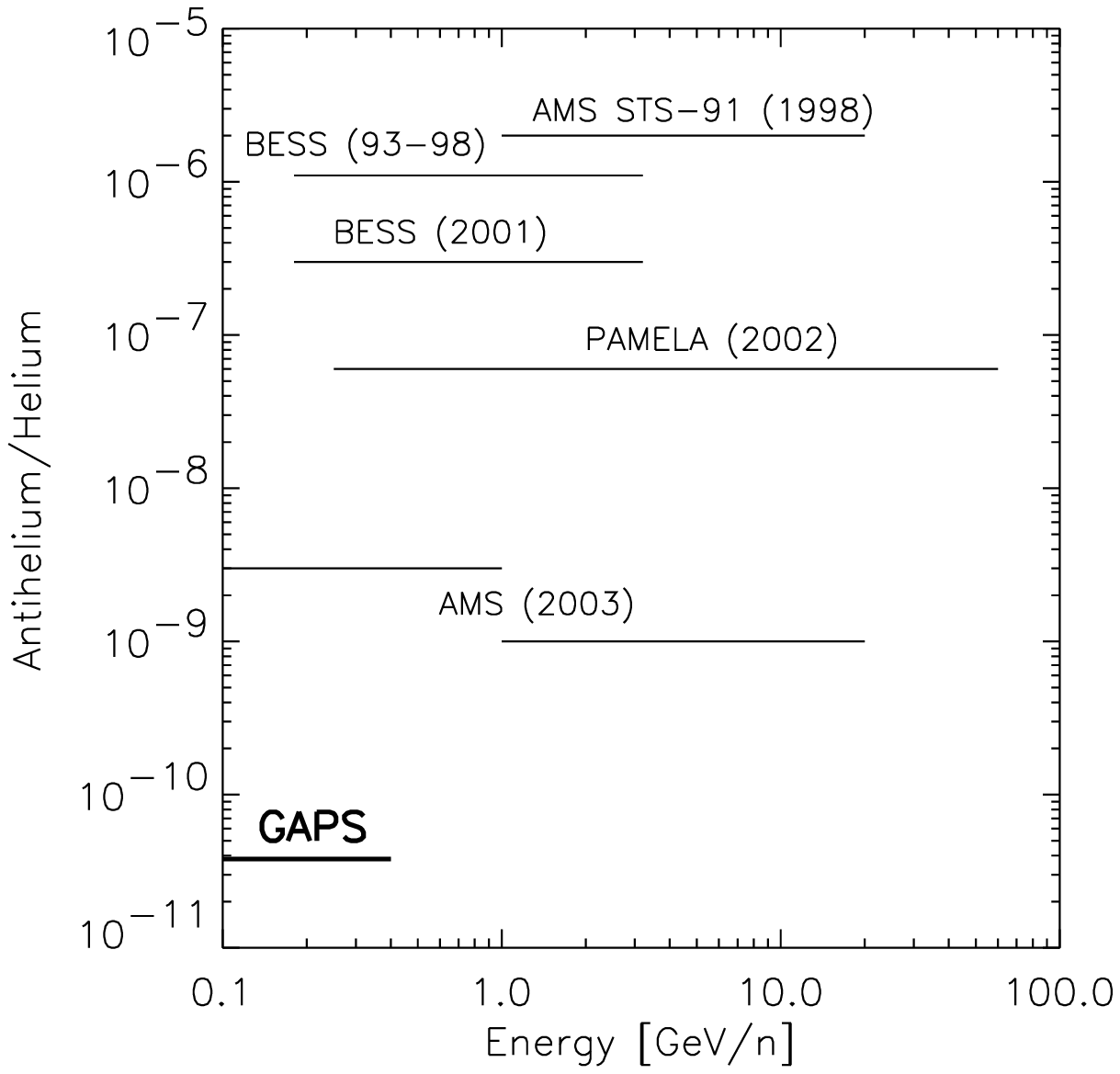}
\end{figure}

Previous measurements  set an upper limit on the ratio
$\bar{He}/He$  of   $10^{-6}$ \citep{saeki98, battiston98}. The AMS/ISS experiment will be able to set
upper limits of $\bar{He}/He$ of $10^{-9}$ with 95\% confidence \citep{battiston99}. In figure \ref{helimit}, we  present the sensitivities of \hebar
detection for our satellite-based antideuteron search.  We obtained  a 
sensitivity  of $\sim  1.5\times10^{-9}$m$^{-2}$sr$^{-1}$GeV$^{-1}$sec$^{-1}$. This translates  to   a  $\bar{He}/He$
of $\sim 3.8\times10^{-11}$  with  95\% confidence, assuming   the  He   flux
in  the GAPS energy band   is
$1.6\times10^2$  m$^{-2}$sr$^{-1}$(GeV/n)$^{-1}$sec$^{-1}$. Our limit is
$\sim100$ times better than the AMS/ISS mission. Our increased sensitivity to 
the $\bar{He}/He$ ratio is partly due to operating at lower energies where the
He flux is more copious, and partly due to greatly increased grasp over
AMS. There are (potential) complications in interpreting the ``real''
improvement in sensitivity versus AMS because the leakage of \hebar into our
galaxy is energy-dependent due to galactic wind and magnetic field effects
\citep{streitmatter96}. However given that the uncertainties in the theory, it
is not meaningful at this time to try to correct for such effects. 



\section{Summary}

We have presented a novel scheme for detecting antimatter which has high
sensitivity, especially at lower energies where conventional techniques are
inefficient. It also provides an unambiguous signal for the presence of
antimatter. Current effort will focus on development of more accurate models of
background in the GAPS detector and development of effective strategies for
using the available information from the detector matrix to reject this
background. Ultimately a prototype will be built and tested. 

\acknowledgements{We  are  grateful  to  Jaesub  Hong  and  Jason  Koglin  for
stimulating discussions and useful comments. This work was partially supported
by NASA grant NAG5-7737.}


\begin{thebibliography}{49}
\expandafter\ifx\csname natexlab\endcsname\relax\def\natexlab#1{#1}\fi

\bibitem[{{Adamo} {et~al.}(1993)}]{adamo93}
{Adamo, A., et al.} 1993, Phys. Rev. A, 47, 4517

\bibitem[{{Akerib}(2000)}]{akerib00}
{Akerib}, D. 2000, in 19th International Conference on Neutrino Physics and
  Astrophysics

\bibitem[{{Andr{\' e}s} {et~al.}(1999)}]{amanda99}
{Andr{\' e}s, E.~C., et al.} 1999, Nuclear Physics B Proceedings
  Supplements, 77, 474

\bibitem[{{Aschenauer} {et~al.}(1995)}]{aschenauer95}
{Aschenauer, E.~C., et al.} 1995, \pra, 51, 1965

\bibitem[{{Bacher} {et~al.}(1988){Bacher}, {Bl{\"u}m}, {Gotta}, {Heitlinger},
  {Schneider}, {Missimer}, {Simons}, \& {Elsener}}]{bacher88}
{Bacher}, R., {Bl{\"u}m}, P., {Gotta}, D., {Heitlinger}, K., {Schneider}, M.,
  {Missimer}, J., {Simons}, L.~M., \& {Elsener}, K. 1988, Phys. Rev. A, 38,
  4395

\bibitem[{{Baltz} \& {Gondolo}(2001)}]{baltz01}
{Baltz}, E. \& {Gondolo}, P. 2001, Phys. Rev. Lett., 86, 5004

\bibitem[{{Baltz} \& {Edsj{\" o}}(1999)}]{baltz99}
{Baltz}, E.~A. \& {Edsj{\" o}}, J. 1999, \prd, 59, 05bd7+

\bibitem[{{Barkas} {et~al.}(1963){Barkas}, {Dyer}, \& {Heckman}}]{barkas63}
{Barkas}, W.~H., {Dyer}, J.~N., \& {Heckman}, H.~H. 1963, \prl, 11, 26

\bibitem[{{Battiston}(1998{\natexlab{a}})}]{battiston99}
{Battiston}, R. 1998{\natexlab{a}}, Nuclear Physics B Proceedings Supplements,
  Volume 65, Issue 1-3, p.\ 19-26., 65, 19

\bibitem[{{Battiston}(1998{\natexlab{b}})}]{battiston98}
{Battiston}, R. 1998{\natexlab{b}}, in Dark matter : proceedings of DM97, 1st
  Italian conference on dark matter, Trieste, December 9-11, 1997 / editor,
  Paolo Salucci. Firenze, Italy : Studio Editoriale Fiorentino, c1998, p. 39.,
  39

\bibitem[{{Beck} {et~al.}(1993){Beck}, {Wilets}, \& {Alberg}}]{beck93}
{Beck}, W.~A., {Wilets}, L., \& {Alberg}, M.~A. 1993, \pra, 48, 2779

\bibitem[{{Belotsky} {et~al.}(2000){Belotsky}, {Golubkov}, {Khlopov},
  {Konoplich}, \& {Sakharov}}]{belotsky00}
{Belotsky}, K.~M., {Golubkov}, Y.~A., {Khlopov}, M.~Y., {Konoplich}, R.~V., \&
  {Sakharov}, A.~S. 2000, Physics of Atomic Nuclei, Volume 63, Issue 2,
  February 2000, pp.233-239, 63, 233

\bibitem[{{Berger} {et~al.}(1999){Berger}, {Coursey}, \& {Zucker}}]{pstar}
{Berger}, M.~J., {Coursey}, J.~S., \& {Zucker}, M.~A. 1999, ESTAR, PSTAR, and
  ASTAR: Computer Programs for Calculating Stopping-Power and Range Tables for
  Electrons, Protons, and Helium Ions (version 1.21), National Institute of
  Standards and Technology, Gaithersburg, MD.

\bibitem[{{Bergstr{\"o}m} {et~al.}(1999){Bergstr{\"o}m}, {Edsj{\"o}}, \&
  {Ullio}}]{bergstrom99}
{Bergstr{\"o}m}, L., {Edsj{\"o}}, J., \& {Ullio}, P. 1999, ApJ, 526, 215

\bibitem[{{Bergstr{\" o}m} {et~al.}(1998){Bergstr{\" o}m}, {Edsj{\" o}}, \&
  {Gondolo}}]{bergstrom98_1}
{Bergstr{\" o}m}, L., {Edsj{\" o}}, J., \& {Gondolo}, P. 1998, \prd, 58, 1945f+

\bibitem[{{Bergstr{\" o}m} \& {Gondolo}(1996)}]{bergstrom96}
{Bergstr{\" o}m}, L. \& {Gondolo}, P. 1996, Astroparticle Physics, 5, 263

\bibitem[{{Bertin} {et~al.}(1996)}]{obelix96}
{Bertin, A., et al.} 1996, Phys. Lett. B, 369, 77

\bibitem[{{Bottino} {et~al.}(1998){Bottino}, {Donato}, {Fornengo}, \&
  {Salati}}]{bottino98}
{Bottino}, A., {Donato}, F., {Fornengo}, N., \& {Salati}, P. 1998, \prd, 58,
  123503

\bibitem[{{Brown} {et~al.}(2001)}]{brown01}
{Brown, H.~N., et al.} 2001, Phys. Rev. Lett., 86, 2227

\bibitem[{{Br{\"u}ckner} {et al.}(1990)}]{bruckner90}
{Br{\"u}ckner, W., et al.} 1990, Z. Phys. A, 335, 217

\bibitem[{{Cohen}(2000)}]{cohen00}
{Cohen}, J.~S. 2000, \pra, 62, 057f0+

\bibitem[{{Day} {et~al.}(1959){Day}, {Snow}, \& {Sucher}}]{day59}
{Day}, T.~B., {Snow}, G.~A., \& {Sucher}, J. 1959, Phys. Rev. Lett., 3, 61

\bibitem[{{Donato} {et~al.}(2000){Donato}, {Fornengo}, \& {Salati}}]{donato00}
{Donato}, F., {Fornengo}, N., \& {Salati}, P. 2000, Phys. Rev. D., 62, 043003

\bibitem[{{Donato} {et~al.}(2001){Donato}, {Maurin}, {Salati}, {Barrau},
  {Boudoul}, \& {Taillet}}]{donato01}
{Donato}, F., {Maurin}, D., {Salati}, P., {Barrau}, A., {Boudoul}, G., \&
  {Taillet}, R. 2001, preprint (astro-ph/0103150)

\bibitem[{{Edsj{\" o}}(1997)}]{edsjo97_2}
{Edsj{\" o}}, J. 1997, PhD thesis, , Uppsala Univ.\ (preprint hep-ph/9704384),
  (1997)

\bibitem[{{Edsj{\" o}} \& {Gondolo}(1997)}]{edsjo97_1}
{Edsj{\" o}}, J. \& {Gondolo}, P. 1997, \prd, 56, 1879

\bibitem[{{Ferrell}(1960)}]{ferrell60}
{Ferrell}, R. 1960, Phys. Rev. Lett., 4, 425

\bibitem[{{Gondolo}(2000)}]{gondolo00}
{Gondolo}, P. 2000, in 19th International Conference on Neutrino Physics and
  Astrophysics

\bibitem[{{Harrison} {et~al.}(2001)}]{harrison01}
{Harrison, F.~A., et al.} 2001, Nuclear Instruments and Methods A, in press

\bibitem[{{Hartmann}(1990)}]{hartmann90}
{Hartmann}, F.~J. 1990, Electromagnetic cascade and chemistry of exotic atoms,
  127

\bibitem[{{Hayano} {et~al.}(1994)}]{hayano94}
{Hayano, R.~S., et al.} 1994, Phys. Rev. Lett., 73, 1485

\bibitem[{{Hubbell} \& {Seltzer}(1997)}]{xcom}
{Hubbell}, J.~H. \& {Seltzer}, S.~M. 1997, Tables of X-Ray Mass Attenuation
  Coefficients and Mass Energy-Absorption Coefficients (version 1.02), National
  Institute of Standards and Technology, Gaithersburg, MD.

\bibitem[{{Jacot-Guillarmod} {et~al.}(1988){Jacot-Guillarmod}, {Bienz},
  {Boschung}, {Piller}, {Schaller}, {Schellenberg}, {Schneuwly}, \&
  {Siradovic}}]{jg88}
{Jacot-Guillarmod}, R., {Bienz}, F., {Boschung}, M., {Piller}, C., {Schaller},
  L.~A., {Schellenberg}, L., {Schneuwly}, H., \& {Siradovic}, D. 1988, Phys.
  Rev. A, 37, 3795

\bibitem[{{Jungman} \& {Kamionkowski}(1994)}]{jungman94}
{Jungman}, G. \& {Kamionkowski}, M. 1994, Phys. Rev. D., 49, 2316

\bibitem[{{Jungman} {et~al.}(1996){Jungman}, {Kamionkowski}, \&
  {Griest}}]{jungman96}
{Jungman}, G., {Kamionkowski}, M., \& {Griest}, K. 1996, Phys. Rep., 267, 195

\bibitem[{{Keck} {et~al.}(2001)}]{keck01}
{Keck, J.~W., et al.} 2001, ApJ, submitted

\bibitem[{{Kurki-Suonio} \& {Sihvola}(2000)}]{kurki-suonio00}
{Kurki-Suonio}, H. \& {Sihvola}, E. 2000, Phys. Rev. Lett., 84, 3756

\bibitem[{{Leo} \& {Haase}(1994)}]{leo90}
{Leo}, W.~R. \& {Haase}, D.~G. 1994, Techniques for Nuclear and Particle
  Physics Experiments (2nd revised edition. ; Springer-Verlag Berlin Heidelberg
  New York.)

\bibitem[{{MacGibbon} \& {Carr}(1991)}]{macgibbon91}
{MacGibbon}, J.~H. \& {Carr}, B.~J. 1991, \apj, 371, 447

\bibitem[{{Maki} {et~al.}(1996){Maki}, {Mitsui}, \& {Orito}}]{maki96}
{Maki}, K., {Mitsui}, T., \& {Orito}, S. 1996, Phys. Rev. Lett., 76, 3474

\bibitem[{{Mandic} {et~al.}(2001){Mandic}, {Baltz}, \& {Gondolo}}]{mandic01}
{Mandic}, V., {Baltz}, E., \& {Gondolo}, P. 2001, in preparation

\bibitem[{{Nakamura} {et~al.}(1994)}]{nakamura94}
{Nakamura, S.~N., et al.} 1994, \pra, 49, 4457

\bibitem[{{Reifenr{\"o}ther} \& {Klempt}(1989)}]{reifenrother89}
{Reifenr{\"o}ther}, G. \& {Klempt}, E. 1989, Nuclear Physics A, 503, 885

\bibitem[{{Ryufuku} {et~al.}(1980){Ryufuku}, {Sasaki}, \&
  {Watanabe}}]{ryufuku80}
{Ryufuku}, H., {Sasaki}, K., \& {Watanabe}, T. 1980, Phys. Rev. A., 21, 745

\bibitem[{{Saeki} {et~al.}(1998)}]{saeki98}
{Saeki, T., et al.} 1998, Phys. Lett. B, 422, 319

\bibitem[{{Simon} {et~al.}(1998){Simon}, {Molnar}, \& {Roesler}}]{simon98}
{Simon}, M., {Molnar}, A., \& {Roesler}, S. 1998, \apj, 499, 250

\bibitem[{{Streitmatter}(1996)}]{streitmatter96}
{Streitmatter}, R.~E. 1996, Nuovo Cimento, 19, 835

\bibitem[{{Wells} {et~al.}(1999){Wells}, {Moiseev}, \& {Ormes}}]{wells99}
{Wells}, J.~D., {Moiseev}, A., \& {Ormes}, J.~F. 1999, ApJ, 518, 570

\bibitem[{{Zombeck}(1982)}]{zombeck82}
{Zombeck}, M.~V. 1982, Handbook of space astronomy and astrophysics (Cambridge,
  Cambridge University Press, 1982, 331 p.)

\end{thebibliography}


\begin{deluxetable}{ccccc}
\tablewidth{0pt}
\tablecaption{Ladder transition energies [keV] for the antiproton. \label{e_pbar}}
\tablehead{Transition & $Z=7$ & $Z=8$ & $Z=10$ & $Z=18$}
\startdata
$2\rightarrow{1}$ & 857.7   &   1130.  &    1787. & 5927.\\
$3\rightarrow{2}$ & 158.8   &   209.2  &    330.9 & 1097.\\
$4\rightarrow{3}$ & \bf{55.59}   &   \bf{73.22}  &    \bf{115.8} & 384.2\\
$5\rightarrow{4} $ & \bf{25.73}  &    \bf{33.89}  &    \bf{53.60} & 177.8\\
$6\rightarrow{5}$ & 13.98   &   18.41   &  \bf{29.12} & \bf{96.60}\\
$7\rightarrow{6}$ & 8.427    &  11.10    &  17.56 & \bf{58.25}\\
$8\rightarrow{7}$ & 5.470    & 7.204    &  11.40 & 37.81\\
$9\rightarrow{8}$ & 3.750   &   4.939  &   7.813 & 25.92\\
\enddata
\tablecomments{Numbers in bold faces are energies of the candidate photons observable by  GAPS.}
\end{deluxetable}

\begin{deluxetable}{ccccc}
\tablewidth{0pt}
\tablecaption{Ladder transition energies [keV] for the antideuteron.\label{e_dbar}}
\tablehead{Transition & $Z=7$ & $Z=8$ & $Z=10$ & $Z=18$}
\startdata
$3\rightarrow{2}$ & 297.8   &   395.1  &   632.0 & 2143.\\
$4\rightarrow{3}$ & \bf{104.2}   &  \bf{138.3}  &    221.2 &750.1 \\
$5\rightarrow{4} $ & \bf{48.24}  &  \bf{64.01}   & \bf{102.4}   &347.2 \\
$6\rightarrow{5}$ & \bf{26.21}   & \bf{34.77}     & \bf{55.61}  &188.6 \\
$7\rightarrow{6}$ & 15.80   & 20.97    &\bf{33.53}   & \bf{113.7} \\
$8\rightarrow{7}$ & 10.26   & 13.61   & 21.76 & \bf{73.81}\\
$9\rightarrow{8}$ & 7.031  &  9.330  & 14.92  & \bf{50.60} \\
$10\rightarrow{9}$ &  5.030   & 6.673  & 10.67 & 36.20\\
$11\rightarrow{10}$ &  3.721   & 4.938  & 7.897 & 26.78\\
\enddata
\end{deluxetable}

\begin{deluxetable}{ccccc}
\tablewidth{0pt}
\tablecaption{Ladder transition energies [keV] for antihelium.\label{e_hebar}}
\tablehead{Transition & $Z=7$ & $Z=8$ & $Z=10$ & $Z=18$}
\startdata
$6\rightarrow{5}$ & 186.4    &  250.4   &   408.1 & 1440.\\
$7\rightarrow{6}$ & \bf{112.4}    &  151.0   &   246.1 & 868.4\\
$8\rightarrow{7}$ & \bf{72.93}    &  \bf{97.98}   &   159.7 & 563.6\\
$9\rightarrow{8}$ & \bf{50.00}    &  \bf{67.17}   &   \bf{109.5} & 386.4\\
$10\rightarrow{9}$ & \bf{35.77}   &   \bf{48.05}   &   \bf{78.32} & 276.4\\
$11\rightarrow{10}$ & 26.46  &    \bf{35.55}   &   \bf{57.95} & 204.5\\
$12\rightarrow{11}$ & 20.13   &   27.04   &   \bf{44.07} & 155.5\\
$13\rightarrow{12}$ & 15.66   &   21.04   &   34.30 & \bf{121.0}\\
$14\rightarrow{13}$ & 12.42   &   16.70  &   27.22 & \bf{96.04}\\
$15\rightarrow{14}$ & 10.03   &   13.47   &   21.96 & \bf{77.48}\\
$16\rightarrow{15}$ & 8.206  & 11.02 & 17.97 & \bf{63.41}\\
\enddata
\end{deluxetable}

\begin{deluxetable}{ccc}
\tablewidth{0pt}
\tablecaption{Configuration of the high latitude mission for detection of the antideuteron and antihelium.\label{dbar_mission}}
\tablehead{ Parameter }
\startdata
Latitude 			& 70$^\circ$N\\
Total size $L$ [m]	 	& 	5 \\
Weight [ton] 			& $\le5$\\
Overall column density [gcm$^{-2}$] 	& 5 \\
Number of cells 		& 27 \\
Gas element 			& Nitrogen \\
Peak effective grasp [m$^2$sr] for $\bar{D}$ ($\bar{He}$) & 9.0 (11.1)\\
Energy band [GeV/n] 		& $0.1-0.4$ \\
\enddata
\tablecomments{The peak effective grasps are for the lowest energy
channel. Band pass of other channels are shifted to high energy by the degrader.}
\end{deluxetable}

\begin{deluxetable}{cccc}
\tablewidth{0pt}
\tablecaption{Comparison of the sensitivity of primary antideuteron 
detection. Observation time is 3 years.\label{dbar_table}}
\tablehead{Observation & $I_{\bar{D}}$ [m$^{-2}$sr$^{-1}$GeV$^{-1}$sec$^{-1}$]
& Energy band [GeV/n]}
\startdata
GAPS on HLM 	& $2.6\times10^{-9}$ 	& $0.1-0.4$ \\
AMS on ISS\tablenotemark{a} 	& $4.8\times10^{-8}$	& $0.1-2.7$
\enddata
\tablenotetext{a}{Solar minimum. It is $3.2\times10^{-8}$ at
solar maximum.}
\end{deluxetable}

\begin{deluxetable}{ccc}
\tablewidth{0pt}
\tablecaption{Estimated number of antiprotons measured by 1 year interplanetary probe.\label{interplanetary} }
\tablehead{Source & $40-60$ MeV & $100-120$ MeV}
\startdata
Secondary (no p-p) & 20	&  30	\\ 
Neutralino annihilation & 60 &  100		\\
Primordial black hole evaporation  & 280	& 400
\enddata
\tablecomments{The effective grasp is 65 cm$^2$sr for each channel.}
\end{deluxetable}


\begin{figure}
\epsscale{0.5}
\caption{Sensitivity of current/planned underground experiments for neutralino
detection for various SUSY models \citep{bergstrom96,edsjo97_1,edsjo97_2,bergstrom98_1,baltz99,mandic01}. The green circled SUSY models are those
allowed if the anomalous muon g--2 result is correct \citep{baltz01}.\label{sensitivity1}}
\plotone{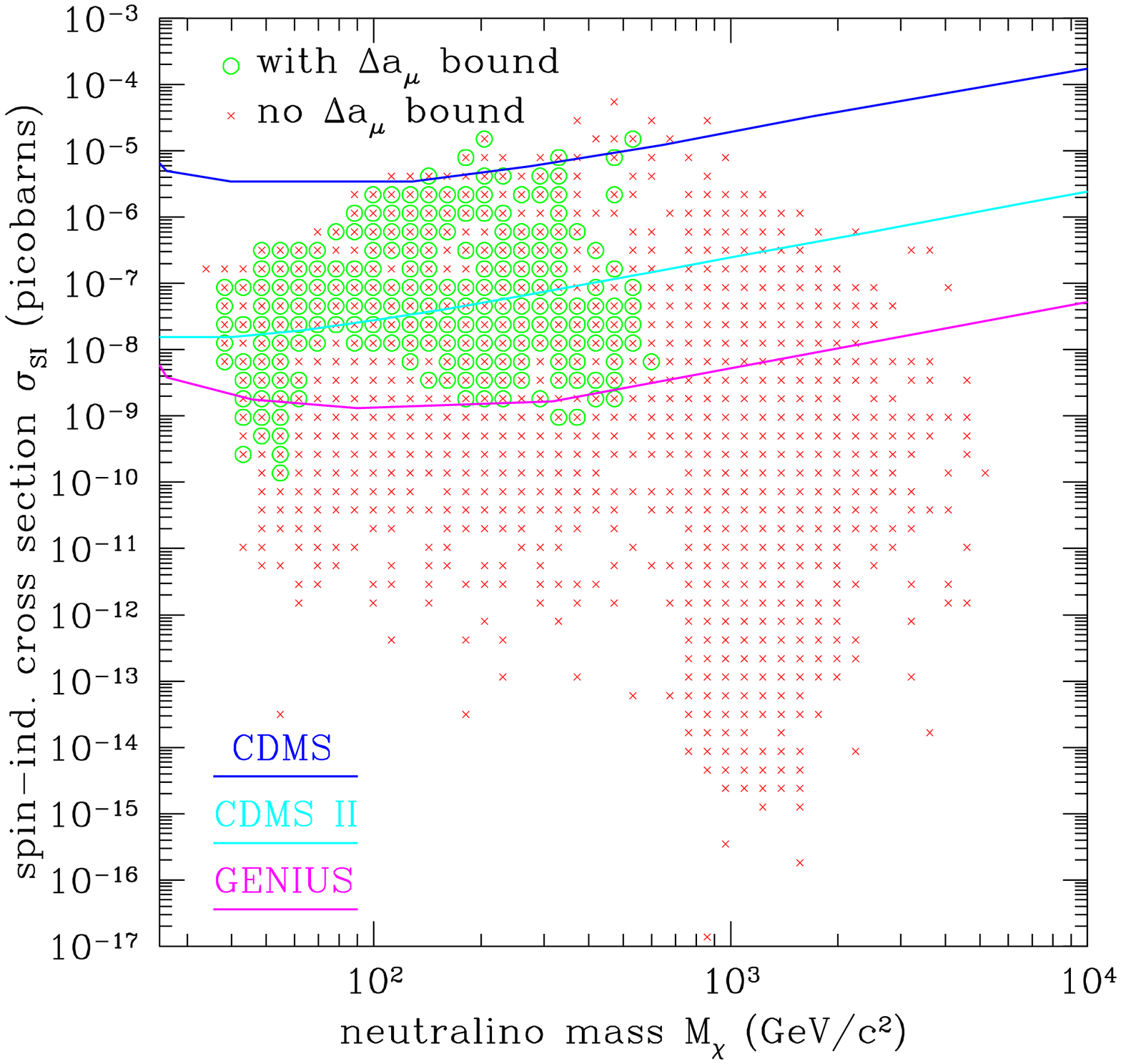}
\end{figure}

\begin{figure}
\epsscale{0.5}
\caption{Sensitivity of a MIDEX class GAPS implementation for neutralino
detection shown with the same set of models as in figure
\ref{sensitivity1}. The antideuteron flux is obtained by rescaling the
solar-modulated antiproton flux of \citet{bergstrom99} by a factor $10^{-4}$ \citep{donato00}. \label{sensitivity2}}
\plotone{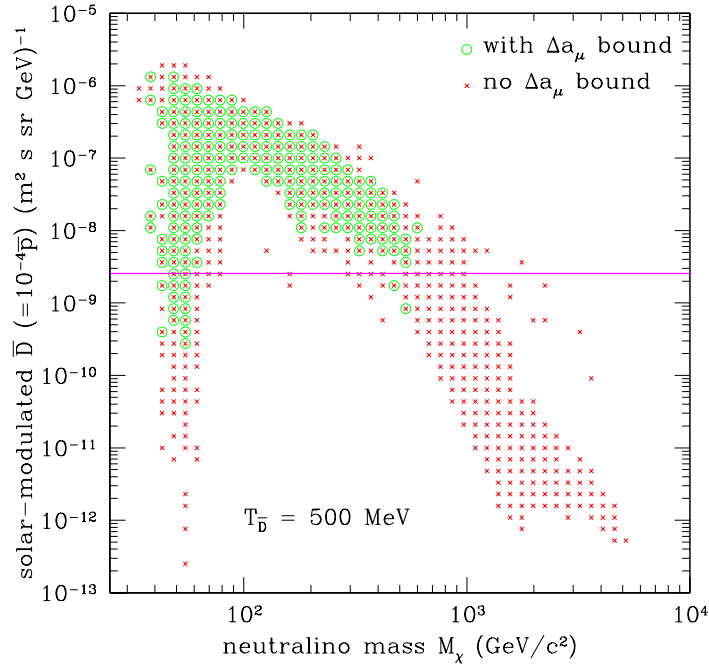}
\end{figure}

\end{document}